\documentclass[
reprint,
superscriptaddress,
bibnotes,
nofootinbib,
prmaterials,
amsmath,amssymb,
aps,
]{revtex4-2}

\usepackage{lipsum}
\usepackage{graphicx}
\usepackage{dcolumn}
\usepackage{bm}
\usepackage{hyperref}
\usepackage[version=4]{mhchem}    
\usepackage{csquotes}
\hypersetup{colorlinks=true,linkcolor=red,citecolor=red}
\usepackage{float}
\usepackage{xspace}
\usepackage{physics}
\usepackage[normalem]{ulem}
\usepackage[dvipsnames]{xcolor}

\usepackage{bbding} 

\usepackage{prettyref}
\newcommand{\pref}[1]{\prettyref{#1}}
\newrefformat{fig}{Fig.~\ref{#1}}
\newrefformat{tab}{Table~\ref{#1}}
\newrefformat{sec}{Sec.~\ref{#1}}
\newrefformat{subsec}{Sec.~\ref{#1}}
\newrefformat{app}{App.~\ref{#1}}
\newrefformat{eqn}{Eq.~(\ref{#1})}

\usepackage[dvipsnames]{xcolor}

\newcommand{\lamno}{LaMnO$_3$\xspace}
\newcommand{\lunio}{LuNiO$_3$\xspace}

\def\QE{\textsc{Quantum ESPRESSO}\,}

\newcommand{\response}[1]{\textcolor{black}{#1}}

\newcommand{\editor}[2]{%
  \expandafter\newcommand\csname #1note\endcsname[1]{%
    \textcolor{#2}{(\textbf{#1:} ##1)}}%
  \expandafter\newcommand\csname #1\endcsname[1]{%
    \textcolor{#2}{##1}}%
  \expandafter\newcommand\csname #1cancel\endcsname[1]{%
    \textcolor{#2}{\sout{##1}}}%
  \expandafter\newcommand\csname #1change\endcsname[2]{%
    \textcolor{#2}{\sout{##1} ##2}}%
  \newenvironment{#1text}{\color{#2}}{\color{black}}
}

\editor{IT}{red}

\begin{document}

\title{Explicit demonstration of the equivalence between DFT+$U$ and the Hartree-Fock limit of DFT+DMFT}

\author{Alberto Carta} \email{alberto.carta@mat.ethz.ch}
\affiliation{Materials Theory, ETH Z\"urich, Wolfgang-Pauli-Strasse 27, 8093 Z\"urich, Switzerland}
\author{Iurii Timrov} \email{ iurii.timrov@psi.ch}
\affiliation{PSI Center for Scientific Computing,
Theory, and Data, 5232 Villigen PSI, Switzerland}
\author{Peter Mlkvik}
\affiliation{Materials Theory, ETH Z\"urich, Wolfgang-Pauli-Strasse 27, 8093 Z\"urich, Switzerland}
\author{Alexander Hampel}
\affiliation{Center for Computational Quantum Physics, Flatiron Institute, 162 5th Avenue, New York, NY 10010, USA}
\author{Claude Ederer} \email{edererc@ethz.ch}
\affiliation{Materials Theory, ETH Z\"urich, Wolfgang-Pauli-Strasse 27, 8093 Z\"urich, Switzerland}

\date{\today}

\begin{abstract}
Several methods have been developed to improve the predictions of density functional theory (DFT) in the case of strongly correlated electron systems. 
Out of these approaches, DFT+$U$, which corresponds to a static treatment of the local interaction, and DFT combined with dynamical mean field theory (DFT+DMFT), which considers local fluctuations, have both proven incredibly valuable in tackling the description of materials with strong local electron-electron interactions.
While it is in principle known that the Hartree-Fock (HF) limit of the DFT+DMFT approach should recover DFT+$U$, demonstrating this equivalence in practice is challenging, due to the very different ways in which the two approaches are generally implemented.
In this work, we introduce a way to perform DFT+$U$ calculations in \QE using Wannier functions as calculated by Wannier90, which allows us to use the same Hubbard projector functions both in DFT+$U$ and in DFT+DMFT.
We benchmark these DFT+$U$ calculations against DFT+DMFT calculations where the DMFT impurity problem is solved within the HF approximation. Considering a number of prototypical materials including NiO, MnO, LaMnO$_3$, and LuNiO$_3$, we establish the sameness of the two approaches.
Finally, we showcase the versatility of our approach by going beyond the commonly used atomic orbital-like projectors by performing DFT+$U$ calculations for VO$_2$ using a special set of bond-centered Wannier functions. 

\end{abstract}

\maketitle


\section{\label{sec:Intro}Introduction}

Hubbard-corrected density functional theory (DFT), in particular the DFT+$U$ approach~\cite{Anisimov1991}, but also the combination of DFT with dynamical mean-field theory (DMFT)~\cite{Georges1996, Anisimov1997}, have emerged as powerful and widely-used computational tools to study transition-metal (TM) compounds or $f$ electron systems~\cite{Anisimov/Aryasetiawan/Lichtenstein:1997, Kotliar/Vollhardt:2004, Held2007, Kunes_et_al:2010, Himmetoglu2014, Paul/Birol:2019}.
In these materials, the strong local Coulomb interaction felt by the more localized electrons can be comparable or greater than the band-width, and a treatment using standard semi-local DFT exchange-correlation (xc) functionals based on the local density approximation (LDA) or the generalized gradient approximation (GGA) can lead to quantitatively or qualitatively wrong results. Antiferromagnetic oxides and, more generally, Mott insulators are prime examples. For these materials, LDA/GGA can severely underestimate the band-gap or even predict a metallic ground state, whereas DFT+$U$ (or DFT+DMFT) can correctly reproduce the band-gap as well as other ground-state properties~\cite{Anisimov/Aryasetiawan/Lichtenstein:1997, Kotliar/Vollhardt:2004, Held2007, Kunes_et_al:2010, Himmetoglu2014, Paul/Birol:2019, KirchnerHall:2021}. 

Within DFT+$U$ and DFT+DMFT, the local (screened) electron-electron interaction, usually denoted by $U$, is treated more explicitly for a subset of the electronic degrees of freedom, the \emph{correlated subspace} or \emph{Hubbard manifold}, which typically corresponds to the $d$ or $f$ electrons. This is achieved by adding the local interaction within the correlated subspace on top of the standard semi-local DFT xc functional, while simultaneously correcting for a potential double-counting. Further extensions that also incorporate explicit inter-site interactions are also used~\cite{LeiriaCampoJr/Cococcioni:2010}. 

The main conceptual difference between the DFT+$U$ and the DFT+DMFT method is the level of approximation used to handle the local interaction. The DFT+$U$ method essentially corresponds to a \emph{static} (i.e., time-independent) Hartree-Fock \response{approximation}\footnote{\response{Note that we use the term ``Hartree-Fock approximation'' in the general sense of a single Slater determinant approximation to the (local) electron-electron interaction,
and not to refer to the specific first principles ``Hartree-Fock method'', which considers the general (non-local) interaction and determines optimal single particle orbitals from a variational principle.
}} 
of the local interaction terms, and thus retains the usual mapping on the auxiliary non-interacting, i.e., uncorrelated, Kohn-Sham (KS) electrons used within DFT, but now with an additional, in general orbital- and spin-dependent potential that depends on the local density matrix and the strength of the screened local interaction.\footnote{Note that the KS potential of course has an xc contribution and as such does also contain correlation effects. However, considered as auxiliary non-interacting particles, the KS electrons are completely uncorrelated and can be described by a single Slater determinant.}
The DFT+DMFT approach, albeit also neglecting explicit non-local correlation effects, consists in a \emph{dynamical} mean-field approximation, which incorporates a frequency-dependent self-energy to capture the true many-body nature of the local correlations, thereby offering a more advanced treatment of the electron-electron interactions.

It has been shown that the DFT, DFT+$U$, and DFT+DMFT approaches can all be understood within a unifying framework based on the Baym-Kadanoff functional~\cite{Kotliar2006}, which is a functional of the full many-body Green's function of the system and is closely related to the grand canonical potential. 
Within this framework, the defining equations for the different methods can be obtained by applying different approximations to the Baym-Kadanoff functional, which then also correspond to choosing different components of the full Green's function as central variables of the resulting theory. In particular, DFT+$U$ can be obtained by considering the electronic density and the local density matrix of the correlated sites as independent variables in the Baym-Kadanoff functional, while in DFT+DMFT one chooses the electronic density combined with the full local Green's function for each correlated site as the fundamental variables~\cite{Kotliar2006}.   

In this theoretical framework, the DFT+$U$ method can also be obtained as the static Hartree-Fock limit to the more general DFT+DMFT approximation. Thus, in this limit, the two approaches become formally identical. Nevertheless, in practice, this equivalence has never been explicitly demonstrated. This is due to numerous technical differences in the way the two methods are typically implemented~\cite{Dang_et_al:2014}.
These differences can include different choices of the correlated subspace and the corresponding basis orbitals, different parameterizations of the local interaction Hamiltonian, whether and how a potential magnetic or orbital symmetry breaking is included, the possible neglect of charge self-consistency within DFT+DMFT, and finally the fact that DFT+$U$ is formulated in an effective single-particle KS framework, while DFT+DMFT uses a many-body Green's function formalism. More details are discussed in \pref{sec:Methods}.

Here, we  explicitly demonstrate that indeed \emph{quantitatively identical results}, essentially within numerical accuracy, can be obtained within DFT+$U$ and DFT+DMFT in the Hartree-Fock limit [from now on referred to as DFT+DMFT(HF)], provided that all technical aspects are treated in an analogous way within the two computational workflows. 
This is accomplished by enabling the use of projections onto (potentially maximally localized) Wannier functions (WFs) to apply the +$U$ correction within the \QE (QE) distribution~\cite{Giannozzi2009, Giannozzi:2017, Giannozzi:2020}, thereby ensuring a consistent definition of the correlated subspace across both approaches. Correspondingly, a Hartree-Fock solver \cite{TRIQS_Hartree_Fock}  is employed to treat the auxiliary local impurity problem within our DMFT framework~\cite{Beck2022}.

We begin in \pref{sec:Methods} by highlighting the similarities and potential differences in the implementation of the two methods, and outline the necessary steps that need to be taken to ensure a consistent comparison. 
Moreover, we introduce a new implementation enabling the use of WFs constructed using Wannier90~\cite{Mostofi_et_al:2014, Pizzi2020}, as DFT+$U$ projectors in QE to represent the correlated subspace.
To benchmark the two approaches against each other, in \pref{sec:Results} we then perform both DFT+$U$ and DFT+DMFT(HF) calculations on several systems of increasing complexity (local magnetic moments in NiO and MnO, orbital order in \lamno, charge disproportionation and structural breathing mode in \lunio), thereby also comparing to DFT+$U$ calculations performed with the L\"owdin-orthonormalized atomic orbitals that are already available as DFT+$U$ projectors in QE~\cite{Timrov:2020b}. 

Apart from filling the gap in the literature of explicitly demonstrating the numerical equivalence between DFT+$U$ and DFT+DMFT(HF), the flexibility of our newly implemented computational tools also allows us to go beyond the use of typical atomic-orbital-like projectors. To highlight this by an example, in \pref{subsec:VO2}, we present DFT+$U$-type calculations for the coupled structural and metal-insulator transition in VO$_2$ using a basis of bond-centered WFs as Hubbard projectors~\cite{Mlkvik_et_al:2024}. Using these unconventional basis orbitals allows for an improved description of the low-temperature insulating phase compared to standard single-site DFT+$U$ performed with atomic orbitals.\\

\section{\label{sec:Methods}Theory and Methods}

Both in the case of DFT+$U$ and DFT+DMFT, the total Hilbert space is separated into two parts: the \emph{correlated subspace} for which the local electron-electron interaction is treated explicitly, and the rest of the electronic degrees of freedom which are treated purely using a standard (semi-) local DFT xc functional. This can be expressed by a Hamiltonian of the type: 
\begin{equation}
    \label{eqn:general_H}
    H =  H^0_{\text{KS}} + H_{\text{int}} - H_{\text{DC}} \quad .
\end{equation}
Here, $H^0_{\text{KS}}$ represents the effective single-particle KS term which contains the kinetic energy of the KS electrons, the full Hartree interaction, and the xc contributions stemming from the particular choice of the underlying functional. The local interaction term, $H_{\text{int}}$, describes the local screened electron-electron interaction restricted to the correlated subspace, and the final term, $H_{\text{DC}}$, represents a double counting (DC) correction. The role of the latter is to cancel out the part of the local interaction in the correlated subspace that is already taken into account in $H^0_\text{KS}$.
We note that the ideal choice of the double counting correction is (to some extent) still an open problem of DFT+$U$/DFT+DMFT~\cite{Karolak_et_al:2010,Park2014}. However, the corresponding discussion is outside of the scope of this work. Our particular choice of double counting correction is defined in \pref{subsec:Comp_det}, \pref{eqn:E_FLL}.

\subsection{\label{subsec:interaction} Definition of the interaction term}  

The most general local two-particle electron-electron interaction Hamiltonian can be written as (see, e.g., \cite{Martin/Reining/Ceperley:2016}):
\begin{align}
    \label{eqn:hint}
   &H_{\text{int}} = \frac{1}{2} \sum_{ \{m\} \sigma \sigma'} \mathcal{U}_{m m' m'' m'''} \,  c^{\dagger  \sigma}_{m} c^{\dagger  \sigma'}_{m''} c^{\sigma'}_{m'''} c^{ \sigma}_{m'} \quad ,
\end{align}
where $\{m\}, \sigma, \sigma'$ indicate different orbital and spin characters. Note that here and in the following, we usually absorb a potential atomic site index into $\{m\}$, unless it is crucial for a correct understanding.  The interaction tensor $\mathcal{U}$ is defined as:
\begin{align}
    \label{eqn:Vee}
    &\mathcal{U}_{m m' m'' m'''} = \bra{m m''} V_\text{ee}\ket{m' m'''}  \nonumber \\
    &=  \int d\mathbf{r} \int d \mathbf{r'} \phi^*_{ m }(\mathbf{r})\phi^*_{ m''}(\mathbf{r'})
    V_\text{ee}(\mathbf{r},\mathbf{r'})
    \phi_{ m' }(\mathbf{r})\phi_{ m'''}(\mathbf{r'}) \quad .
\end{align}
If \pref{eqn:hint} is used in \pref{eqn:general_H} to represent the local interaction term within the DFT+$U$ or DFT+DMFT context, then the annihilation and creation operators, $c^\sigma_m$ and $c^{\dagger \sigma}_m$, and the orbitals $\phi_m(\mathbf{r})$ entering \pref{eqn:hint} and \pref{eqn:Vee} correspond to the basis orbitals that define the correlated subspace, and $V_\text{ee}$ represents the partially screened Coulomb interaction within this subspace.

\subsection{\label{subsec:dftu_dmft_projectors} Projector functions in DFT+$U$ and DFT+DMFT}

The first major difference between specific implementations of the two approaches arises from the choice of the correlated subspace and from the choice of 
the localized orbitals $\phi_m(\mathbf{r})$, i.e., the projector functions, that are used to describe the part of the Hilbert space where the local interaction is treated explicitly by $H_\text{int}$.
It should be noted that this choice is solely based on physical and chemical intuition and an optimal choice will in general depend on the system under consideration. 
While in most cases atomic-like $d$ or $f$ orbitals are a natural choice, in other cases more unconventional orbitals, e.g., a molecular-orbital-like basis set, might be appropriate~\cite{Solovyev:2008, Kovacik_et_al:2012, Ferber_et_al:2014, Grytsiuk_et_al:2024, Mlkvik_et_al:2024}.
Furthermore, one should note that even conventional atomic-like orbitals are generally not uniquely defined within a solid and different definitions are typically employed, differing, e.g., in terms of their radial parts, truncation, orthogonalization, or other details.

In virtually all DFT+$U$ implementations, the Hubbard manifold is defined implicitly by the specific choice of projector functions, and thus differences between different implementations arise mainly due to different definitions of these projectors (see below). On the other hand, in the context of DFT+DMFT calculations, it is also rather common to explicitly select the correlated subspace by considering only a reduced set of bands within a certain energy window, containing, e.g., only bands with dominant TM $d$ character, or an even smaller subset corresponding only to the $t_{2g}$ or $e_g$ sub-shells. The Hubbard projectors are then defined completely within that group of bands.
A related idea has recently been explored also at the DFT+$U$ level, by introducing a separate $+U$ correction on the $t_{2g}$ and $e_g$ states~\cite{Macke:2024}.

In DFT+$U$ implementations, the corresponding projectors are usually hard-coded and depend on the specific code used for the simulations~\cite{Wang2016}. For example, QE~\cite{Giannozzi2009} implements the possibility to use non-orthogonalized and orthonormalized atomic-like orbitals~\cite{Timrov:2020b} (see Appendix~\ref{app:NOA_and_LOAO}), the ``Vienna Ab-initio Simulation Package'' (VASP)~\cite{Kresse1996} projects the crystal wavefunctions onto local orbitals defined in the augmentation regions of the projector augmented wave (PAW) method~\cite{Bengone2000}, while in full-potential linear augmented plane wave (FLAPW) codes, the partial waves calculated within the muffin-tin spheres are a natural choice~\cite{Schick1999}. 

In DFT+DMFT, similar code-specific projectors are also employed~\cite{Pourovskii_et_al:2007, Haule2010}. However, many implementations are also based on potentially more flexible WFs~\cite{Marzari_et_al:2012}, either maximally localized ones or defined by projections of local orbitals on the Bloch functions within a certain energy window and subsequent orthogonalization~\cite{Anisimov_et_al:2005, Lechermann2006, Amadon_et_al:2008, Trimarchi_et_al:2008, Beck2022}, as outlined in Appendix~\ref{app:WFs}. 
In fact, there are also some instances where WFs have been used as projectors for DFT+$U$ calculations~\cite{Fabris:2005, Fabris_et_al:2005, Korotin2012, Korotin2014, Novoselov:2015, Ting:2023}.

To conduct a quantitative comparison between the DFT+DMFT(HF) and DFT+$U$ approaches, we must ensure that identical projectors are used in both cases.

\subsection{\label{subsec:wannier_projectors} Definition of Wannier projectors}

The DFT KS states within a periodic solid are expressed in terms of one-electron Bloch states $\psi_{\mathbf{k} \nu} (\mathbf{r})$ which depend on the $\mathbf{k}$-point within the Brillouin zone (BZ) of the lattice and on the band index $\nu$.
One can then construct WFs, $\mathrm{W}_{m \mathbf{R}}(\mathbf{r})$~\cite{Marzari_et_al:2012, Pizzi2020}, from a subset of Bloch functions within a certain \emph{energy window}:
\begin{equation}
| \mathrm{w}_{m \mathbf{k}} \rangle = \sum_{\nu} |\psi_{\mathbf{k} \nu} \,\rangle U_{\nu m \mathbf{k}} \ ,
\label{eqn:Wannier_generic}
\end{equation}
and Fourier transforming:
\begin{equation}
    \label{eqn:wann_state}
    |\mathrm{W}_{m \mathbf{R}} \rangle = V \int_{\text{BZ}} \frac{d \mathbf{k}}{(2 \pi)^3} e^{-i \mathbf{k \cdot R}} \, |\mathrm{w}_{m \mathbf{k}} \rangle \ .
\end{equation}
Thereby, $\mathbf{R}$ denotes the Bravais lattice vector of the unit cell where the WF is localized and $m$ specifies all other characteristics, such as orbital character or potentially a specific atomic site or spin character, $V$ is the unit cell volume, and the summation over $\nu$ in \pref{eqn:Wannier_generic} runs over all bands within the chosen energy window.

The gauge freedom in the definition of the Bloch states is inherited by the WFs in the form of the matrices $U_{\nu m \mathbf{k}}$. 
Thus, the resulting WFs are not uniquely defined, and different schemes correspond to different choices of the gauge matrices $U_{\nu m \mathbf{k}}$. The Wannier90 code~\cite{Pizzi2020} provides the possibility to obtain maximally localized WFs (MLWFs), defined by choosing $U_{\nu m \mathbf{k}}$ such that the sum of the quadratic spreads of the WFs is minimized~\cite{Marzari_et_al:2012}.
However, a simpler case corresponds to building WFs without performing the spread minimization and instead just projecting the KS wavefunctions onto Bloch sums of atomic orbitals and using the L\"owdin orthonormalization method~\cite{Lowdin:1950}. This type of WFs, which are often used in DFT+DMFT, are typically used as initial WFs in Wannier90 and are described in more detail in Appendix~\ref{app:WFs}.

Apart from the gauge matrices $U_{\nu m \mathbf{k}}$, the shape and spatial extension of the WFs also depends strongly on the chosen energy window in \pref{eqn:Wannier_generic}. For instance, in TM oxides, constructing WFs confined to the states near the Fermi level with dominant TM $d$ character results in more extended orbitals with significant oxygen $p$ contributions (so called oxygen tails), reflecting the strong $p$-$d$ hybridization typically present in these systems~\cite{Lechermann2006, Scaramucci_et_al:2015}.
In contrast, a wider energy window encompassing both TM $d$ and oxygen $p$ dominated bands typically yields more localized WFs resembling atomic-like $d$ orbitals centered on the TM sites [examples are shown in \pref{fig:Nio_big}(c)-(g)], and simultaneously also atomic-like $p$ orbitals centered on the oxygen sites.
It has been shown, that different choices of the energy window can in some cases lead to significant differences in the description of the physics in the corresponding DFT+DMFT calculations~\cite{Parragh2013}.

We note that, despite their non-uniqueness, WFs depend solely on the underlying Bloch states, the definition of a specific energy window, and the criterion to construct $U_{\nu m \mathbf{k}}$. By specifically utilizing maximally localized Wannier projectors, DFT+$U$ and DFT+DMFT calculations can in principle be made quantitatively comparable across different DFT codes, offering a natural way to standardize these results.

\subsection{\label{subsec:U_correction} Hartree-Fock approximation and the equivalence between DFT+$U$ and DFT+DMFT(HF)}

The Hartree-Fock approximation consists in applying the following mean-field decoupling to \pref{eqn:hint} (see, e.g., \cite{Martin/Reining/Ceperley:2016}):
\begin{align}
&\langle \langle c^{\dagger  \sigma}_{m} c^{\dagger  \sigma'}_{m''} c^{ \sigma'}_{m'''} c^{ \sigma}_{m'} \rangle \rangle  \nonumber \\
&\sim \langle \langle c^{\dagger  \sigma}_{m} c^{ \sigma}_{m'} \rangle \rangle \langle\langle c^{\dagger  \sigma'}_{m''} c^{ \sigma'}_{m'''} \rangle\rangle - \langle\langle c^{\dagger  \sigma}_{m} c^{ \sigma'}_{m'''} \rangle\rangle \langle\langle c^{\dagger  \sigma'}_{m''} c^{ \sigma}_{m'} \rangle\rangle \nonumber \\
&:= n^{ \sigma \sigma}_{m m'}n^{ \sigma' \sigma'}_{m'' m'''}-n^{ \sigma \sigma'}_{m m'''}n^{ \sigma' \sigma}_{m'' m'} \ ,
    \label{eqn:decoupling}
\end{align}
where $\langle \langle \cdot \rangle \rangle$ indicates the corresponding expectation values.
This results in the following interaction energy $E_\text{int} = \langle\langle H_\text{int} \rangle\rangle$:
\begin{align}
   \label{eqn:E_int_methods}
   E_{\text{int}}  &\simeq  \frac{1}{2}\sum_{ \{m\} \sigma } \mathcal{U}_{m m' m'' m'''} \, n^{ \sigma}_{m m'} n^{ \bar{\sigma}}_{m'' m'''} \nonumber \\
   &+ \frac{1}{2} \big( \mathcal{U}_{m m' m'' m'''} - \mathcal{U}_{m m' m''' m''} \big) \, n^{ \sigma}_{m m'} n^{ \sigma}_{m'' m'''} \,,
\end{align}
where $\bar{\sigma}=-\sigma$ and we assumed collinear spin order without spin-orbit coupling so that the spin-off-diagonal elements of the density matrix can be discarded, allowing us to condense the notation $n^{ \sigma\sigma}_{m m'}$ to $n^{ \sigma}_{m m'}$.
Up to here, the same expression is used to approximate the local interaction energy in both DFT+$U$ and DFT+DMFT(HF). The difference arises in the way that the correction is implemented in practice.

In DFT+$U$, the Hubbard correction results in a local and in general orbital- and spin-dependent contribution to the KS potential $\mathcal{V}_\text{KS}^\sigma$ for sites $I$ in unit cell $\mathbf{R}$: 
\begin{equation}
   \label{eqn:V_dftu}
  \mathcal{V}_{\text{KS}}^\sigma = \mathcal{V}^0_{\text{KS}} + \sum_{I,\mathbf{R}} \mathcal{V}^{\sigma}_{C, I \mathbf{R}} \,.
\end{equation}
Here, $\mathcal{V}^0_{\text{KS}}$ is the usual ionic, Hartree, and xc contribution to the KS potential, and $\mathcal{V}^\sigma_{C}$ is the local Hubbard potential defined as:
\begin{equation}
\label{eqn:V-Hub}
\mathcal{V}^\sigma_C = \sum_{m m'}(\mathcal{V}_C)^{\sigma}_{m m'} \ket{\mathrm{W}_{m}} \bra{\mathrm{W}_{m'}} \quad ,
\end{equation}
where we dropped the site indices again for better readability.
Note that here we take $\mathcal{V}^0_{\text{KS}}$ as spin-independent, since in all our calculations we treat spin-polarization only at the $+U$ or +DMFT level and evaluate the DFT xc functional from the spin-averaged density (see \pref{subsec:Comp_det} for more details).
The $\ket{\mathrm{W}_{m}}$ in \pref{eqn:V-Hub} are the Hubbard projectors corresponding to the local orbitals, which in our case are WFs as defined in Eq.~\eqref{eqn:wann_state} (with the unit-cell and site index absorbed into $m$ for ease of notation).
We also note that, as indicated in \pref{eqn:V-Hub}, within this work we generally use spin-independent WFs that are constructed from an initial non-spin-polarized DFT calculation, or, in the case of our DFT+DMFT(HF) calculations, are obtained from spin-independent KS states corresponding to $\mathcal{V}_\text{KS}^0$ only.
$(\mathcal{V}_C)^{ \sigma}_{m m'}$ is obtained as~\cite{Kotliar2006}:
\begin{equation}
   \label{eqn:V_dftu_1}
  (\mathcal{V}_C)^{\sigma}_{m m'} = \frac{ \delta \langle\langle H_{\text{int}} - H_{\text{DC}} \rangle \rangle }{\delta n^{\sigma}_{m m'}} \quad ,
\end{equation}
and the occupation matrices $n^{\sigma}_{m m'}$ are defined in terms of the local Hubbard projectors as:
\begin{equation}
    \label{eqn:occupations_dftu}
     n^{\sigma}_{m m'} = \sum_{\mathbf{k} \nu} f(\varepsilon_{\mathbf{k} \nu}^\sigma) \braket{\mathrm{W}_{m}}{\psi^\sigma_{\mathbf{k} \nu}}\braket{\psi^\sigma_{\mathbf{k} \nu}}{\mathrm{W}_{m'}}.
\end{equation}
Here, $f(\varepsilon_{\mathbf{k} \nu}^\sigma)$ denotes the occupation of the KS eigenstate $\ket{\psi_{\mathbf{k} \nu}^\sigma}$ with eigenvalue $\varepsilon_{\mathbf{k} \nu}^\sigma$.
The KS equations with the effective single-particle potential from \pref{eqn:V_dftu} are solved iteratively until self-consistency with respect to the KS wavefunctions, $\psi_{\mathbf{k}\nu}^\sigma(\mathbf{r})$, the corresponding charge density, and the orbital density matrix, \pref{eqn:occupations_dftu}, is achieved.

In contrast, in DFT+DMFT, the theory is formulated in terms of Green's functions~\cite{Georges1996, Kotliar2006}, where one treats the explicit local interaction term by mapping each correlated site to an effective impurity model.
In doing so, the $\mathbf{k}$-dependence of the self-energy in the local basis describing the correlated subspace is neglected and only the dependence on frequency $\omega$ is considered, $\Sigma^\sigma_{mm'}(\omega, \mathbf{k}) \rightarrow \Sigma^\sigma_{mm'}(\omega)$.
The DMFT self-consistency condition requires that each impurity Green's function
is equal to the corresponding local component, $G_{C}(\omega)$, of the total Green's function, $G(\omega, \mathbf{k})$, of the full lattice problem.

If we apply the Hartree-Fock approximation to the solution of the effective impurity model, the local self-energy becomes also frequency-independent, $\Sigma^\sigma_{mm'}(\omega) \medspace {\stackrel{\mathcal{\normalfont\mbox{\scriptsize{HF}}}}{\longrightarrow}} \medspace \Sigma^\sigma_{mm'}$, so that the self-energy ultimately amounts to an in general orbital- and spin-dependent static potential shift on the local orbitals. More explicitly, we can express $\Sigma$ in the local basis as:
\begin{align}
   \label{eqn:sigma_hf}
  &\Sigma^{\sigma}_{m m'}(\omega) =  \frac{\delta \langle \langle H_{\text{int}} - H_{\text{DC}} \rangle \rangle  }{\delta G_C(\omega)^{\sigma}_{m m'}} \medspace
  {\stackrel{\mathcal{\normalfont\mbox{\scriptsize{HF}}}}{\longrightarrow}} \medspace
  \frac{\delta \langle \langle H_{\text{int}} - H_{\text{DC}} \rangle \rangle  }{\delta n^{\sigma}_{m m'}} \quad .
\end{align}
Thus, in this approximation the local self energy becomes identical to the orbital-dependent potential contribution ${\cal V_C}$ used in DFT+$U$ [{\it cf.} \pref{eqn:V_dftu_1}], i.e., both quantities  depend on the local occupation matrix $n^{\sigma}_{mm'}$ in exactly the same way.
Next, we verify that both DFT+DMFT(HF) and DFT$+U$ also result in equivalent expressions of the Green's functions in terms of the underlying KS wavefunctions.

In DFT+DMFT, the KS equations defining the single particle part of the problem do not explicitly contain the local interaction (Hubbard) term which instead gets incorporated at the DMFT level, i.e, it is added to the Green's function in the form of a local self-energy via the Dyson equation.
The lattice Green's function within DFT+DMFT is thus expressed as follows (see, e.g., \cite{Beck2022}):
\begin{align}
   \label{eqn:DMFT_G}
   G(\omega, \mathbf{k})^{-1} =  
   \omega + i\delta + \mu - \sum_{\alpha} \lambda_{\mathbf{k} \alpha} \ket{\chi_{\mathbf{k} \alpha}}\bra{\chi_{\mathbf{k} \alpha}}  \nonumber \\ -
   \sum _{\alpha \beta m m' \sigma} ( P^\dagger(\mathbf{k}) )_{\alpha m} \,\Sigma^\sigma_{mm'} \, P_{m' \beta}(\mathbf{k}) \ket{\chi_{\mathbf{k}\alpha}} \bra{\chi_{\mathbf{k} \beta}} \, .
\end{align}
Here, $i \delta$ is a small imaginary part,  $\mu$ is the chemical potential, $\lambda_{\mathbf{k}\alpha}$ and $\ket{\chi_{\mathbf{k}\alpha}}$ are the eigenvalues and corresponding eigenstates (with band index $\alpha$) of the KS problem without local interaction and double counting terms, $\Sigma_{mm'}$ is the frequency-independent local self energy from \pref{eqn:sigma_hf}, and the projector matrices $P_{m\alpha}(\mathbf{k}) = \bra{\mathrm{w}_{m \mathbf{k}}}\ket{\chi_{\mathbf{k}\alpha}}$ encode the transformation from the local orbitals to the Bloch basis.

Using the definition of the projector matrices together with \pref{eqn:sigma_hf}, \pref{eqn:V_dftu_1}, and \pref{eqn:V-Hub} (and considering that $\mathcal{V}_C$ is diagonal in all site-related indeces), the DFT+DMFT Green's function can also be written as:
\begin{align}
   \label{eqn:dftu_transformed}
   G(\omega , &\mathbf{k})^{-1} = \omega + i \delta  + \mu  \\ \nonumber
   &- \sum_{ \alpha \beta \sigma} \ket{\chi_{\mathbf{k} \alpha}} \left\{ \lambda_{\mathbf{k} \alpha} \delta_{\alpha\beta} + \bra{\chi_{\mathbf{k} \alpha}} \mathcal{V}^\sigma_{C}  \ket{\chi_{\mathbf{k} \beta}} \right\} \bra{\chi_{\mathbf{k} \beta}} \\ \nonumber
   &= \omega +i \delta + \mu \\ \nonumber
   &- \sum_{\alpha \beta \sigma} \ket{\chi_{\mathbf{k} \alpha}}\bra{\chi_{\mathbf{k} \alpha}} (-\frac{\nabla^2}{2} + \mathcal{V}^0_{\text{KS}} + \mathcal{V}_{C}^\sigma )
   \ket{\chi_{\mathbf{k} \beta}}\bra{\chi_{\mathbf{k} \beta}}
   \quad ,
\end{align}
where we also used $(-\frac{\nabla^2}{2} + \mathcal{V}^0_{\text{KS}}) \ket{\chi_{\mathbf{k} \alpha}} = \lambda_{\mathbf{k} \alpha} \ket{\chi_{\mathbf{k} \alpha}}$. 
Thus, if we now assume that $G$ in \pref{eqn:dftu_transformed} corresponds to a converged fully charge self-consistent solution, then we can alternatively obtain $G$ by diagonalizing the KS equation corresponding to the DFT+$U$ potential from \pref{eqn:V_dftu}, $(-\frac{\nabla^2}{2} + \mathcal{V}^0_{\text{KS}} + \mathcal{V}_{C}^\sigma)\ket{\psi^\sigma_{\mathbf{k}\nu}} = \varepsilon^\sigma_{\mathbf{k}\nu} \ket{\psi^\sigma_{\mathbf{k}\nu}}$, with eigenstates $\ket{\psi^\sigma_{\mathbf{k} \nu}}$ and eigenvalues $\varepsilon^\sigma_{\mathbf{k}\nu}$.
Finally, transforming from the basis of $\chi_{\mathbf{k}\alpha}$ to $\psi_{\mathbf{k}\nu}$ by inserting and removing the corresponding resolutions of identity $1_\mathbf{k} = \sum_{\nu \sigma} \ket{\psi^\sigma_{\mathbf{k} \nu}} \bra{\psi^\sigma_{\mathbf{k} \nu}}$ and $1_\mathbf{k} = \sum_{\alpha \sigma} \ket{\chi_{\mathbf{k} \alpha}} \bra{\chi_{\mathbf{k} \alpha}}$ (where $\ket{\chi_{ \mathbf{k} \alpha}}$ is the same for spin up and down), one obtains:
\begin{equation}
   \label{eqn:dftu_G}
   G(\omega , \mathbf{k})^{-1} =   
   \omega + i \delta  + \mu -  \sum_{\nu\sigma}\varepsilon^\sigma_{\mathbf{k} \nu} \ket{\psi^\sigma_{{\mathbf {k} \nu}}}\bra{\psi^\sigma_{{\mathbf {k} \nu}}} \,.
\end{equation}
One can see that \pref{eqn:dftu_G} is nothing but the Green's function expressed in the KS eigenfunctions of the DFT+$U$ problem, i.e., the solutions of the KS equation with the Hubbard-corrected potential from \pref{eqn:V_dftu}. Since this DFT+$U$ Green's function is formally identical to the DFT+DMFT(HF) Green's function in \pref{eqn:DMFT_G}, with $\ket{\chi_{\mathbf{k}\alpha}}$, $\Sigma_{mm'}^\sigma$, and $\ket{\psi^\sigma_{\mathbf{k}\nu}}$ evaluated from the same charge density and occupation matrices, and considering that in turn both density and local occupation matrices are fully determined by $G(\omega,\mathbf{k})$, this shows that indeed both DFT+$U$ and DFT+DMFT(HF) have the same stationary solution and are therefore fully equivalent to each other.

\subsection{\label{subsec:Comp_det} Computational details}

Our modified version of QE that includes the possibility to use WFs generated by Wannier90 as DFT+$U$ projectors is based on QE version 6.6. This implementation (see Appendix~\ref{app:Interface}) will be publicly distributed with the next offical release of QE~\cite{Giannozzi2009}.

Our DFT+DMFT(HF) calculations are based on solid\_dmft~\cite{Merkel2022}, a computational framework based on the TRIQS library~\cite{Parcollet2015}, using the DFT+DMFT interface described in \cite{Beck2022} and QE version 7.2. We solve the effective impurity problem for each TM ion within the unit cell using the openly available TRIQS Hartree-Fock solver~\cite{TRIQS_Hartree_Fock}. 
Since our DFT+DMFT implementation is based on an imaginary time formulation, we use an inverse temperature of $\beta = 40$\,eV$^{-1}$. 
However, the results show minimal dependence on this temperature choice. 
For consistency, we also apply a Fermi-Dirac smearing with $\beta$ = 40 eV$^{-1}$ for the Brillouin zone integrations in our DFT+$U$ calculations.
Further specifics regarding the choice of structural parameters, $\mathbf{k}$-grid, pseudopotentials, and the specific xc functional used for the different benchmark materials are provided in the supplementary information (SI)~\cite{SI}.

\begin{figure*}
   \centering   \includegraphics[width=0.8\textwidth]{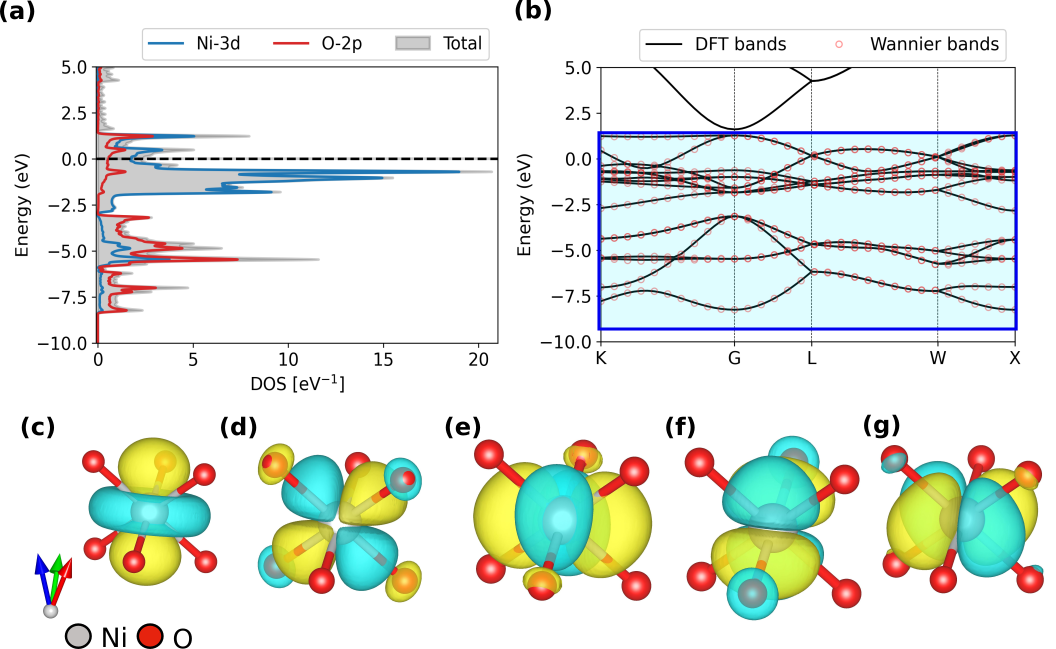}
   \caption{(a) Density of states (DOS) for nonmagnetic NiO, including the projections on the Ni $d$ (blue line) and O $p$ (red line) character. The zero of energy corresponds to the Fermi level. (b) NiO band structure obtained from spin-unpolarized DFT calculations (black lines) together with bands reconstructed from the tight-binding Hamiltonian expressed in the WFs basis (red circles). The blue rectangle indicates the energy window used to construct the WFs, which includes all states with dominant Ni $d$ and O $p$ character. (c)-(g) Plots of the corresponding $d$-like WFs centered on the Ni atom.} 
   \label{fig:Nio_big}
\end{figure*}

To generate the Hubbard projectors for each material, we first perform a non-spin-polarized self-consistent DFT+$U$ calculation with a vanishing Hubbard $U$ correction (see Appendix~\ref{app:Interface}). 
We then construct WFs using Wannier90 (v3.1.0)~\cite{Mostofi_et_al:2014, Pizzi2020}, considering a wide energy window that spans all states with dominant TM $d$ and O $p$ character, as shown in \pref{fig:Nio_big}(a) and \pref{fig:Nio_big}(b) for the case of NiO. Corresponding figures for the other materials employed in this work can be found in the SI~\cite{SI}.
To keep the comparison as close as possible to the projectors based on L\"owdin-orthonormalized atomic orbitals (LOAO) (see Appendix~\ref{app:NOA_and_LOAO}), we refrain from minimizing the overall spread of the WFs and use the initial WFs obtained from projections of the Bloch states within the chosen energy window onto atomic-like orbitals and subsequent orthonormalization (see Appendix~\ref{app:WFs}). 
However, in cases with entangled bands we perform the usual disentanglement procedure to obtain WFs spanning an optimally connected subspace~\cite{Souza/Marzari/Vanderbilt:2001, Mostofi_et_al:2014, Marzari_et_al:2012}.
We also performed test calculations where we minimized the total quadratic spread of the WFs, but we found only small quantitative differences in the results.

In all our calculations, both DFT+DMFT(HF) and DFT+$U$, spin polarization and magnetic order are treated at the level of the correlated subspace only, i.e., the standard DFT xc functional is evaluated only from the spin-averaged charge density, resulting in a spin-independent KS potential ${\cal V}_\text{KS}^0$. This is in analogy with the approach used, e.g., in Refs.~\cite{Chen2016} and \cite{Ryee2018}. 
In practice, this requires a slight modification of the double counting correction (see below) and to zero out the magnetization density in QE before the evaluation of the xc functional. 
Hence, whenever we refer to local magnetic moments, these are calculated from the difference in trace of the spin up and spin down components of the local occupation matrix.

Except for the calculations presented in \pref{subsec:VO2}, we use the usual Slater parametrization, based on the spherical approximation, to represent the local interaction among the TM $d$ orbitals, see, e.g., Ref.~\cite{Liechtenstein1995}, where the matrix elements $\mathcal{U}_{m m' m'' m'''}$ are expressed in terms of Slater parameters, $F^k$ ($k \in \{0, 2, 4\}$), with $U=F^0$, $J = (F^2+F^4)/14$, and $F^4/F^2 = 0.63$. This is hard-coded in QE (so-called \emph{Liechtenstein scheme}) and mirrored in our DFT+DMFT(HF) implementation.
For the DC correction, we use the charge-only fully-localized limit (cFLL)~\cite{Anisimov_et_al:1993, Chen2016, Ryee2018}:
\begin{equation}
    \label{eqn:E_FLL}
    E^{\text{cFLL}}_{\text{DC}} =  \frac{1}{2} U N(N-1)
    - \frac{1}{2} J N (\frac{N}{2}-1) \,,
\end{equation}
where $N$ is the total number of electrons in the correlated shell.
The $U$ and $J$ values used for the different benchmark materials are listed in the corresponding sections. We note that we generally compare results obtained for different projectors using the same values for $U$ and $J$, since here our focus is to analyze potential differences between closely related computational methods. On the other hand, the ``optimal'' values of the Hubbard parameters for a given material of course depend on the specific choice of projectors and generally differ for different projectors.

Finally, we note that there is a small difference in how we handle the construction of the WFs in the two methods. In the DFT+$U$ calculations with Wannier projectors [from now on referred to as DFT+$U$(WF)], we fix the WFs to those computed from the initial self-consistent DFT+$U$ calculation (with $U \approx 0$). In contrast, in the DFT+DMFT(HF) calculations, the WFs are re-calculated at every DFT iteration during the charge-self consistent procedure.
Although this typically does not impact results significantly, we have identified one case in our benchmarks (calculations for LaMnO$_3$ presented in \pref{subsec:LaMnO3}) where this indeed leads to small quantitative discrepancies. This difference can then be removed by recalculating the Hubbard projectors also for the DFT+$U$(WF) case, meaning that the WFs are recalculated from the $U$-corrected ground state (see \pref{subsec:LaMnO3} for a more detailed discussion).

The data used to produce the results of this work is available in the Materials Cloud Archive database~\cite{MaterialsCloudArchive2024}.

\section{\label{sec:Results}Benchmark calculations}

\subsection{\label{subsec:NiO_MnO} Magnetic moments in MnO and NiO}

We start by considering the simple TM monoxides MnO and NiO, characterized by nominal $d^5$ and $d^8$ electronic configurations of the TM elements, respectively.
Both have been extensively investigated in the context of correlated antiferromagnetic insulators and have frequently been used as benchmark materials for the DFT+$U$, DFT+DMFT, and other advanced DFT-based methods, see, e.g., Refs.~\cite{Terakura1984, Anisimov1991, Anisimov_et_al:1993, Tran_et_al:2006, Kunes_et_al:2007, Trimarchi2018, Binci:2024, Bonfa:2024}. 
Both feature a rock-salt structure with a paramagnetic spin configuration above their N\'eel temperatures, while below these temperatures they acquire a small rhombohedral distortion with an antiferromagnetic (so called type II) ground state corresponding to a wave-vector along the [111] direction~\cite{Roth:1958}.
While many previous DFT+$U$ studies have focused on the dependence of the band gap on the Hubbard $U$ parameter, a recent study has also addressed the dependence of the  magnetic moments on the Hund's coupling $J$~\cite{Ryee2018}, which was shown to be sensitive to subtle aspects of different DFT+$U$ flavors. 
This provides an instructive test case for our quantitative comparison of DFT+$U$(WF) and DFT+DMFT(HF).

\begin{figure}
   \centering
   \includegraphics[width=0.9\columnwidth]{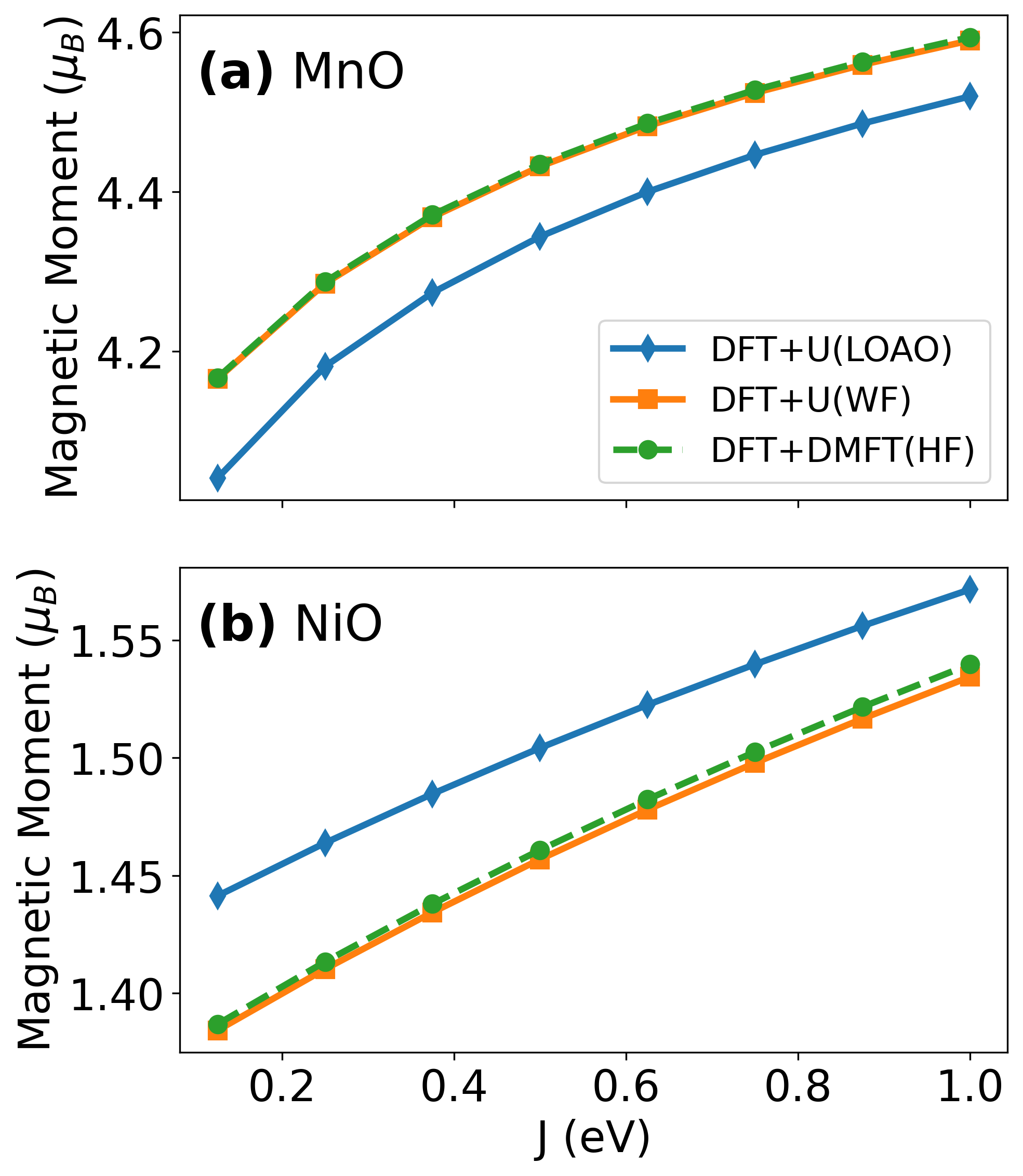}
   \caption{Dependence of the magnetic moments of TM ions for antiferromagnetic compounds (a) MnO (using $U=3$ eV) and (b) NiO (using $U=5$ eV) on the Hund's coupling $J$, obtained from DFT+$U$ calculations with LOAO projectors (blue solid line) or Wannier projectors (orange solid line), and from DFT+DMFT(HF) (green dashed line).}
   \label{fig:NiO_MnO}
\end{figure}

In our calculations, we impose the experimentally observed antiferromagnetic (type II) order and use relaxed structural parameters, where we neglect the small rhombohedral distortions and consider a perfect rock-salt structure (see the SI~\cite{SI}). 
We use a value of $U=3$ eV for the Mn $3d$ states of MnO and $U=5$ eV for the Ni $3d$ states of NiO, following Ref.~\cite{Ryee2018}.

Figures~\ref{fig:NiO_MnO}(a) and \ref{fig:NiO_MnO}(b) show the absolute values of the magnetic moments of the TM ions in MnO and NiO, respectively, as a function of $J$, obtained from DFT+$U$ calculations using LOAO projectors (blue solid line), Wannier projectors (orange solid line), and DFT+DMFT(HF) calculations (green dashed line).
In all cases, the magnetic moment increases with $J$ ({\it cf.} also with Ref.~\cite{Ryee2018}). 
Heuristically, a larger $J$ lowers the energy cost of having several electrons with the same spin on the same site, thereby increasing the magnetic moment. 
Most importantly, we observe that the DFT+DMFT(HF) and DFT+$U$(WF) results are essentially on top of each other. The very small difference noticeable for the case of NiO could be due to the fact that in the DFT+DMFT(HF) calculations the WFs are recomputed at every DFT+DMFT iteration, whereas in the DFT$+U$ case they are calculated only once in the initial self-consistent calculation, as described in \pref{subsec:Comp_det}. This aspect is analyzed in more detail for the case of LaMnO$_3$ in \pref{subsec:LaMnO3}.
On the other hand, there is a clear quantitative difference between DFT+$U$(LOAO) and DFT+$U$(WF)/DFT+DMFT(HF), since these two methods use different Hubbard projectors. 

The essentially numerically identical results for DFT+$U$(WF) and DFT+DMFT(HF) demonstrate that, at least for this case, the two approaches are indeed completely equivalent. In addition, the observation that the difference between DFT+$U$(WF) and DFT+$U$(LOAO) is relatively small, and that the trend with increasing $J$ is very similar, can also be viewed as successful validation of the correctness of the interface between Wannier90 and the DFT+$U$ implementation in QE.

\subsection{\label{subsec:LaMnO3}Orbital order and band gap in LaMnO$_3$}

Next, we consider \lamno, a prototypical material exhibiting orbital order and a cooperative Jahn-Teller distortion below 750\,K~\cite{RodrguezCarvajal1998, tokura2000orbital, Chatterji2003, Pavarini2010, Khomskii:2014}. 
The Jahn-Teller distortion lifts the degeneracy of the $e_g$ states of the Mn$^{3+}$ ion ($d^4$ electronic configuration: $t_{2g}^3 e_g^1$) and stabilizes a staggered orbital ordering along the pseudo-cubic [110] direction (so called C-type order~\cite{Wollan1955}).

\begin{figure}
   \centering
   \includegraphics[width=0.9\columnwidth]{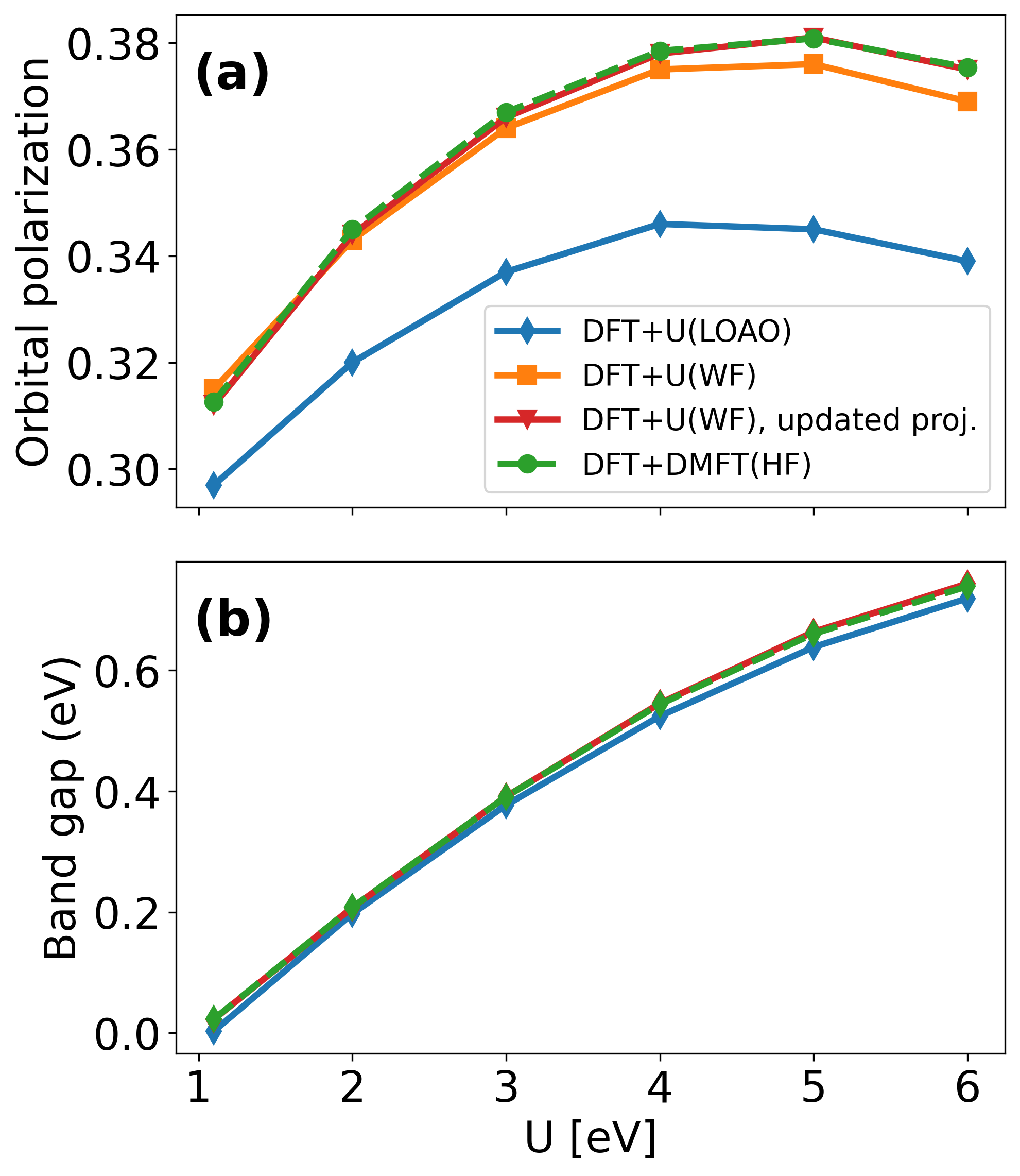}
   \caption{Hubbard $U$ dependence of (a) the orbital polarization and (b) the band gap in \lamno (using $J=1$ eV) for the different methods. We also highlight the effect of recomputing the Wannier projectors at the converged density within DFT+$U$(WF) (\textit{updated proj.}, red line) versus computing them only once at the start. For better visibility in panel (b), we avoid plotting the band gap for the DFT+$U$(WF) one-shot calculations. The corresponding data shows only negligible differences from the other data sets.}
   \label{fig:lamno}
\end{figure}

In our calculations, we use the experimentally determined crystal structure at room temperature~\cite{RodrguezCarvajal1998} with the experimentally observed A-type antiferromagnetic order (wave-vector along the [100] cubic direction)~\cite{Wollan1955}. We systematically vary the Hubbard parameter $U$ for the $3d$ states of Mn, while fixing the Hund's coupling $J$ to a typical value of 1\,eV.
Thereby, the main effect of the Hubbard correction is to enhance the local energy splitting between the more and less occupied $e_g$ states (see, e.g., Ref.~\cite{Kovacik/Ederer:2011}). 

\pref{fig:lamno} illustrates the dependence of both the local orbital polarization and the calculated bandgap on $U$. Here, to define the orbital polarization, we obtain all eigenvalues of the occupation matrix, identify the ones corresponding to the two $e_g$ states, and then take the difference between these two eigenvalues.
We see that already for $U=1$\,eV, the orbital polarization is non-zero (due to the orthorhombic symmetry of the underlying crystal structure), whereas the bandgap is still closed. As $U$ increases, the orbital polarization also increases, albeit rather moderately, reflecting the increased local energy splitting between the two $e_g$ orbitals, until it  eventually reaches a maximum around $U=5$ eV.

Again, there is a clear quantitative difference in the orbital polarization between the DFT+$U$ calculations with the Wannier projectors and the LOAO projectors. In fact, similar to the case shown in \pref{fig:NiO_MnO}(b), there is also a small but noticeable difference between the DFT+DMFT(HF) results and the DFT+$U$(WF) results where the Wannier projectors are only computed once, i.e., after the initial self-consistent calculation (green versus orange lines in Fig.~\ref{fig:lamno}). On the other hand, if the Wannier projectors within the DFT+$U$(WF) approach are re-calculated for the final DFT+$U$ ground state charge density (red line), consistent with the treatment within the DFT+DMFT(HF) method (see~\pref{subsec:Comp_det}), perfect numerical agreement is achieved between the two methods. 

Finally, we observe little dependence of the band gap on the particular choice of projectors, and also the overall qualitative trends are independent of the choice of projectors, consistent with our observations in~\pref{subsec:NiO_MnO}.
This benchmark thus also confirms the general equivalence between DFT+DMFT(HF) and DFT+$U$, but also demonstrates that subtle differences in the workflow can affect the numerical agreement between the two methods.

\subsection{\label{subsec:LuNiO3}Charge disproportionation in LuNiO$_3$}

For our final benchmark, we use \lunio, which is part of the series of rare-earth nickelates that exhibit a metal-to-insulator transition strongly coupled to a structural distortion~\cite{Alonso1999, Alonso_et_al:2000, Medarde2008, GarcaMuoz2009, Catalano2018}. 
Thereby, the nominal Ni$^{3+}$ ions ($d^7$ electronic configuration: $t_{2g}^6 e_g^1$) disproportionate according to $2 \text{Ni}^{3+} \rightarrow \text{Ni}^{2+} + \text{Ni}^{4+}$, with inequivalent Ni cations arranged in a three-dimensional checkerboard pattern. The Ni disproportionation is accompanied by a contraction and expansion of the oxygen octahedra surrounding the nominal Ni$^{4+}$ and Ni$^{2+}$ ions, respectively~\cite{Medarde2009, Catalano2018}. This \emph{breathing mode distortion} is represented by a specific symmetry adapted distortion mode~\cite{Perez-Mato/Orobengoa/Aroyo:2010} with symmetry label $R_1^+$ relative to the ideal cubic perovskite structure~\cite{Balachandran/Rondinelli:2013, Hampel2017}.

Following Ref.~\cite{Hampel2017}, we start from the relaxed structure of \lunio and then vary the amplitude of the $R_1^+$ breathing mode distortion. We impose an A-type antiferromagnetic order as a simpler proxy for the more complex experimentally observed magnetic order ({\it cf.} Refs.~\cite{Hampel2017, Binci:2023}).
We also note that within the atomic $d$-orbital-like basis used in our DFT+DMFT(HF) and DFT+$U$ calculations (for both WF and LOAO), the charge disproportionation is rather subtle and only results in a small difference in the corresponding local charges. In this basis, the disproportionation is better represented within a ligand-hole picture~\cite{Mizokawa/Khomskii/Sawatzky:2000, Johnston_et_al:2014, Varignon_et_al:2017}.
In the following, we therefore quantify the degree of disproportionation by the difference of the magnetic moments on the two inequivalent  Ni sites. 

\begin{figure}
   \centering
   \includegraphics[width=0.9\columnwidth]{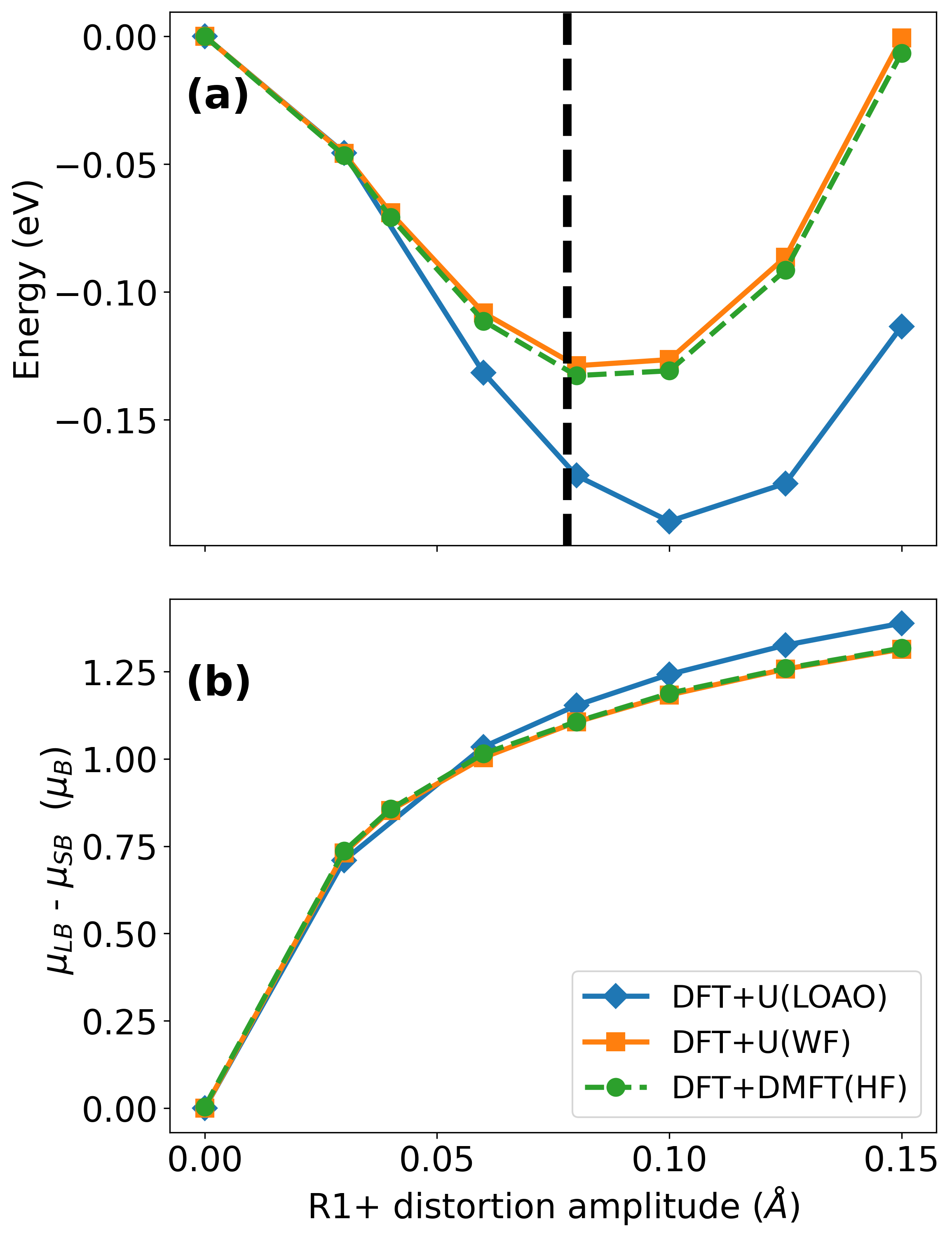}
   \caption{(a) Total energy per LuNiO$_3$ unit cell (4 formula units) relative to the undistorted structure and (b) difference of the magnetic moments of the two inequivalent Ni sites (surrounded by larger, $\mu_{\text{LB}}$, and smaller, $\mu_{\text{SB}}$, octahedra) in LuNiO$_3$ as a function of the $R^+_1$ distortion. Calculations are performed for A-type antiferromagnetic order, $U=5$\,eV, and $J=1$ eV. The vertical dashed line indicates the experimentally observed distortion amplitude \cite{Alonso2001}.}
   \label{fig:lunio_big}
\end{figure}

Figure~\ref{fig:lunio_big}(a) depicts the evolution of the total energy of \lunio as a function of the $R_1^+$ mode, obtained using $U=5$\,eV and $J=1$ eV for the $3d$ states of Ni ({\it cf.} Ref.~\cite{Hampel2017}). 
All three variants of our Hubbard-corrected methods lead to a reduction in the total energy for nonzero distortion, with the DFT+$U$(WF) and DFT+DMFT(HF) results being essentially numerically identical. The small difference can again be ascribed to the fact that in the DFT+$U$(WF) calculations we do not update the Wannier projectors in each iteration step ({\it cf.}~\pref{subsec:LaMnO3}).

Remarkably, the use of the LOAO projectors leads to notable quantitative differences in this case, not only in the depth of the energy minimum but also in its position, which corresponds to a $\sim25$\,\% larger distortion amplitude compared to DFT+$U$ with Wannier projectors. 
Thereby, it should be noted that, while with our choice of interaction parameters the DFT+$U$(WF) and DFT+DMFT(HF) calculations seem to lead to better agreement with the experimental distortion amplitude, this of course depends on the specific choice of $U$ and $J$ and also on the imposed magnetic order. The optimal values of $U$ and $J$ for a given system in turn depend on the specific projectors. To obtain a true ``first principles'' prediction, the $U$ and $J$ parameters should therefore be computed consistently for the specific projectors. However, in the present case, our focus is to analyze differences between computational methods using a consistent set of fixed parameters. Thus, the more favorable agreement with the experimental distortion amplitude in \pref{fig:lunio_big}(a) should not be interpreted as indication that one choice of projectors is superior to another. 

Finally, \pref{fig:lunio_big}(b) illustrates the difference in the magnetic moments between the nominally $\text{Ni}^{2+}$ (long-bond, LB) and $\text{Ni}^{4+}$ (short-bond, SB) sites, $\mu_\text{LB} -\mu_\text{SB}$, computed from the difference of the trace of the local density matrix for each spin component on each site. 
It can be seen that the structural distortion results in a significant difference in these local moments, indicating the electronic disproportionation between inequivalent Ni sites, with a larger (smaller) moment on the LB (SB) site. In this case, the quantitative differences between the three different methods are small, with the DFT+$U$(WF) and DFT+DMFT(HF) data again perfectly aligned.

\section{Perspective: DFT+$U$ calculations with more general Wannier projectors}

In the previous sections, we have successfully demonstrated the numerical equivalence between the DFT+$U$ approach and the Hartree-Fock limit of DFT+DMFT. The crucial point was to ensure that identical Hubbard projector functions are used in both cases to define the correlated subspace. To achieve this, we have implemented the possibility to use general WFs (generated, e.g., by Wannier90~\cite{Pizzi2020}) as DFT+$U$ projectors in QE, as described in Appendix~\ref{app:Interface}. 

To allow for a direct comparison with the LOAO projectors, we have used WFs resembling typical atomic-like orbitals in all our benchmarks. However, the Wannier projectors are generally more flexible (see, e.g., \cite{Fabris:2005, Fabris_et_al:2005, Korotin2012, Korotin2014, Novoselov:2015, Ting:2023} for previous examples of the use of WFs as DFT+$U$ projectors) and also allow us to perform DFT+$U$ calculations based on other, less conventional types of orbitals. Examples for the latter are  molecular-orbital-like basis sets~\cite{Solovyev:2008, Kovacik_et_al:2012, Ferber_et_al:2014, Grytsiuk_et_al:2024, Mlkvik_et_al:2024} or WFs located at defect centers~\cite{SoutoCasares/Spaldin/Ederer:2019, SoutoCasares/Spaldin/Ederer:2021}. In the following we present an example for the use of such more exotic Wannier projectors for the case of VO$_2$.

\subsection{\label{subsec:VO2}Bond-centered orbitals in VO$_2$}

\begin{figure}
\centering
\includegraphics[width=0.9\columnwidth]{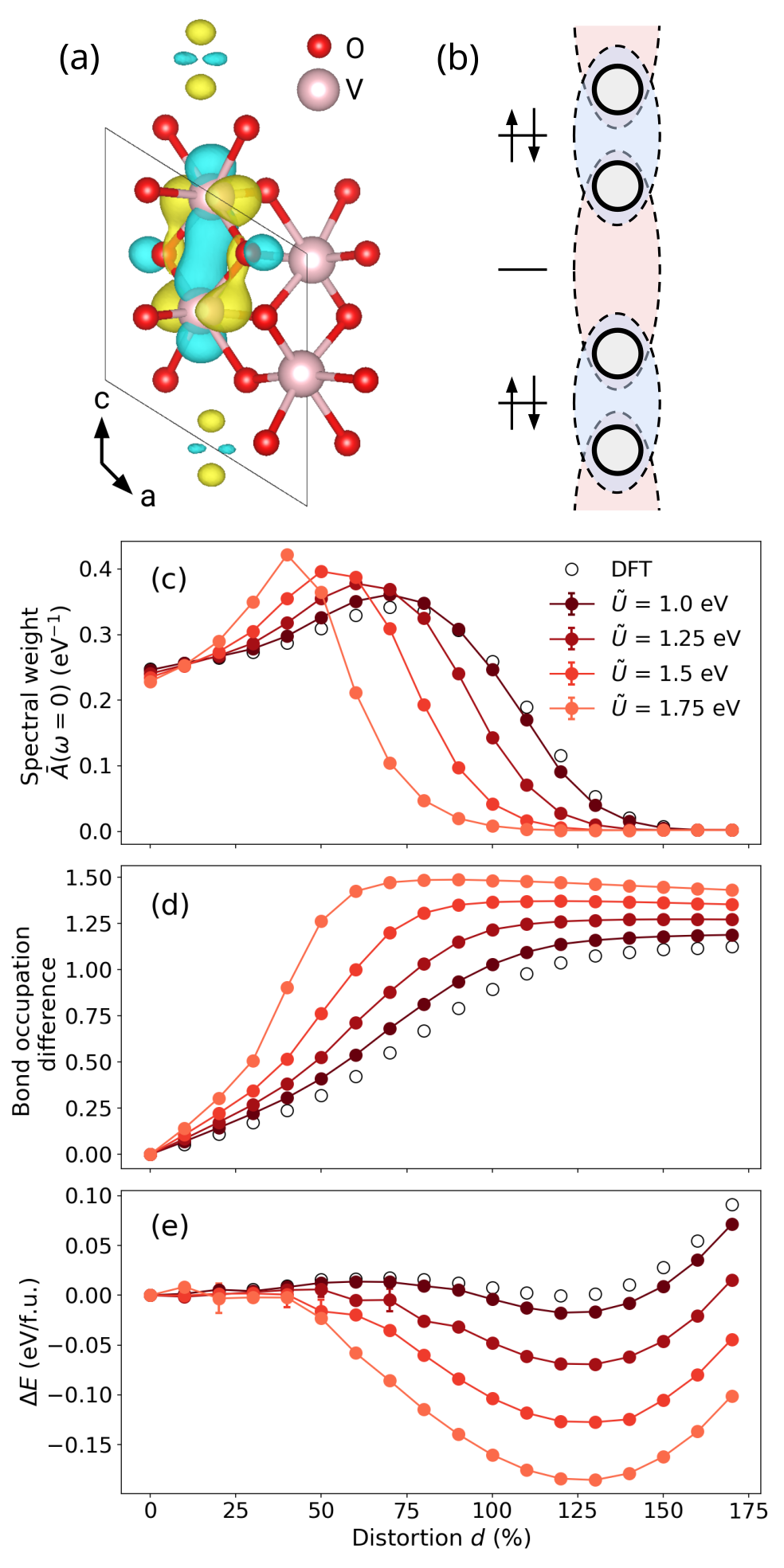}
\caption{(a) Example of a bond-centered WF for the M1 structure of VO$_2$. V (O) atoms are shown in pink (red), and yellow (cyan) isosurfaces show the positive (negative) phase of the orbital. (b) Schematic diagram of singlet formation and bond-charge-disproportionation along V--V chains in VO$_2$. Circles represent V atoms and ovals indicate the bond-centered WFs. (c) Spectral weight at zero frequency, (d) bond occupation difference, and (e) total energy difference $\Delta E$ per formula unit as a function of the structural distortion leading from rutile (0\,\%) to M1 (100\,\%). DFT[+DMFT(HF)] data are shown as empty [filled] markers. The color gradient corresponds to an increasing $\tilde{U}$ value at fixed $\tilde{J}=0.2$\,eV.}
\label{fig:vo2}
\end{figure}

VO$_2$ is another prototypical material undergoing a coupled metal-insulator and structural transition, resulting in the formation of V--V dimers along the $c$ direction of the high temperature rutile structure and a symmetry lowering to a monoclinic (M1) phase below 340\,K \cite{Eyert:2002}. Simultaneously, the magnetic susceptibility drops to nearly zero \cite{Kosuge1967, Pouget_et_al:1974}. The transition mechanism has been debated for many years and is now understood to arise from a combination of Peierls-like singlet formation and Mott-Hubbard correlations~\cite{Goodenough:1971, Zylbersztejn/Mott:1975, Wentzcovitch/Schulz/Allen:1994, Rice/Launois/Pouget:1994, Biermann_et_al:2005, Gatti_et_al:2007, Weber_et_al:2012, Brito_et_al:2016, Najera_et_al:2017, Grandi/Amaricci/Fabrizio:2020}. It is also generally agreed that an accurate DFT-based description of VO$_2$ requires methods beyond standard (semi-) local xc energy functionals~\cite{Eyert:2002, Zhang_et_al:2024}.

However, DFT+$U$ with atom-centered Hubbard projectors does not seem to provide any improvement for VO$_2$~\cite{Mellan_et_al:2019, Stahl/Bredow:2020}. Instead, it leads to a higher energy of the M1 phase compared to the undistorted rutile structure, a strong underestimation of structural dimerization, and an incorrect prediction of magnetic moments in the insulating M1 phase. These deficiencies are related to the importance of inter-site effects that are not well described by standard DFT+$U$ using conventional atom-centered orbitals~\cite{Koch/Manzhos/Chaker:2022, Zhang_et_al:2024}. 

Here, following Ref.~\cite{Mlkvik_et_al:2024}, we perform DFT+DMFT(HF) calculations using Wannier  projectors constructed as bonding combinations of atom-centered $t_{2g}$-type orbitals located on neighboring V atoms. The resulting bond-centered WFs [see an example in Fig.~\ref{fig:vo2}(a)] are constructed between each V--V pair along the rutile $c$ axis, as schematically shown in Fig.~\ref{fig:vo2}(b). Their occupation can be interpreted as ``bond occupation'', i.e., the number of electrons associated to the two vanadium atoms constituting that bond. We then vary the structural distortion in VO$_2$ by linearly interpolating between rutile and M1 structures (see the SI~\cite{SI} for details) and analyze the resulting metal-insulator transition.

In Figs.~\ref{fig:vo2}(c), (d), and (e) we show the spectral weight around zero frequency, $\bar{A}(\omega =0)$, which essentially represents the density of states around the Fermi level, the difference in the bond-centered occupations, and the change in the energy of the system as a function of distortion for different values of $\tilde{U}$ at fixed $\tilde{J}=0.2$\,eV. Note that here, the correlated subspace consists of only three WFs per bond-center and we are using a Kanamori parametrization of the local interaction Hamiltonian (see, e.g., Refs.~\cite{Vaugier/Jiang/Biermann:2012, Georges2013}) instead of the Slater parametrization described in \pref{subsec:interaction} (see the SI~\cite{SI}). We therefore use alternative symbols to represent the intra-orbital interaction, $\tilde{U}$, and the Hund's exchange, $\tilde{J}$.
As a reference, the values of the corresponding interaction parameters computed using the constrained random phase approximation are $\tilde{U}=1.35$\,eV and $\tilde{J}=0.19$\,eV~\cite{Mlkvik_et_al:2024}. We also compare to plain DFT results ($\tilde{U}=\tilde{J}=0$\,eV), shown as empty circles. 

In all cases, one can observe a metal-insulator transition for increasing distortion, indicated by the transition from finite to zero $\bar{A}(\omega =0)$ [\pref{fig:vo2}(c)], accompanied by a pronounced increase of the bond occupation difference between the shorter and the longer V--V bond from zero to around 1.5 [\pref{fig:vo2}(d)]. 
The large difference in the bond occupation represents the electronic dimerization (singlet formation) on the shorter V--V bond, as schematically shown in \pref{fig:vo2}(b). For increasing $\tilde{U}$, the transition moves towards smaller distortion and becomes sharper.
Simultaneously, the energy of the distorted insulating M1 phase is significantly lowered compared to the undistorted metallic rutile structure [\pref{fig:vo2}(e)]. We note that for small $\tilde{U}$ (and for plain DFT), there are two energy minima, one for zero distortion and one around 125\,\% distortion, with the undistorted case in fact being lower in energy within plain DFT. Thus, applying $\tilde{U}$ within the bond-centered basis correctly stabilizes the insulating distorted M1 phase. We note that all our calculations are performed for the spin-degenerate case, i.e., without introducing magnetic moments on the vanadium atoms.

We also note that these results are rather similar to those obtained by applying a static inter-site potential between the V--V pairs along $c$~\cite{Haas_et_al:2024}, and are also consistent with the previous full DFT+DMFT calculations using bond-centered WFs~\cite{Mlkvik_et_al:2024}. This reflects the observation that the inter-site self-energy in the M1 phase becomes essentially frequency-independent, but that this static inter-site effect is necessary to correctly capture the physics of the metal-insulator transition in VO$_2$, as pointed out in earlier works~\cite{Biermann_et_al:2005, Tomczak/Biermann:2007, Tomczak/Aryasetiawan/Biermann:2008, Belozerov_et_al:2012}. The possibility of applying DFT+$U$ on bond-centered WFs hence opens a new way of incorporating these effects in the simple static interaction limit.

\section{Conclusions and Outlook}

In conclusion, we applied the Hartree-Fock approximation to DFT+DMFT to explicitly test its equivalence with the DFT+$U$ method. To enable a truly quantitative comparison between these two approaches, we had to ensure a consistent definition of the correlated subspace by employing WFs as DFT+$U$ projectors within QE. 
The corresponding interface between QE and Wannier90 will be publicly available in the next official release of QE.

Through a series of benchmarks of increasing complexity, we demonstrated that both methods can indeed be brought into perfect agreement, with high numerical precision for various physical observables including magnetic moments, orbital polarization, and equilibrium lattice distortions. Minor numerical discrepancies between the approaches can arise due to subtleties in the way the Wannier projectors are updated (or not) during the self-consistency cycle.

We then showed that the use of WFs allows for a greater flexibility by going beyond standard atomic-orbital-like projectors.  We illustrated this point with the case of VO$_2$, for which we applied the DFT+$U$ correction on an unconventional set of V--V bond-centered orbitals, successfully describing the insulating M1 phase and the metal-insulator transition using only a static mean-field correction.

\response{We point out that due to the Hartree Fock approximation for the local
impurity problem, the ``DFT+DMFT(HF)'' approach results in a purely static
treatment of the local interaction and, as we demonstrate, becomes
completely equivalent to DFT+$U$ (in spite of the ``DMFT'' in our acronym).
Thus, it also inherits the known limitations of the DFT+$U$ method, such
as the inability to describe an insulating state without at least a
local symmetry breaking in cases with degenerate partially occupied
orbitals, e.g., cubic FeO~\cite{Cococcioni:2005, Mazin_Anisimov:1997, Trimarchi2018}.}

\response{
In practice, our work also provides two different publicly available implementations to perform DFT+$U$ calculations with Wannier functions generated by Wannier90, either using our extensions to QE or using the DFT+DMFT(HF) workflow implemented in solid\_dmft~\cite{Merkel2022}. While the QE implementation is more convenient to use for standard calculations employing an atom-centered basis representing the full five-orbital $d$ shell, the solid-DMFT workflow provides more flexibility, in particular for cases with more exotic orbitals such as the bond-centered orbitals in VO$_2$ (see \pref{subsec:VO2}). Ultimately, the choice of which implementation to use is a matter of convenience. Our work shows that the result will not depend on this choice.
}

\response{Finally, w}e also point out that performing DFT+$U$ (or, equivalently, DFT+DMFT in the Hartree-Fock approximation) with WFs in principle allows for a better comparability between DFT+$U$ calculations performed with different codes. Furthermore, using a consistent choice of basis functions for the correlated subspace allows one to gradually incorporate more complexity in the treatment of local interaction effects, and, e.g.,  systematically compare the results of including dynamic versus static correlation effects.  Another interesting possibility is to start by including dynamical correlations in a magnetically ordered state typically treated within DFT+$U$ and then proceed to address the paramagnetic case.

\begin{acknowledgments}
This research was supported by ETH Z\"urich. I.T. acknowledges support by the NCCR MARVEL, a National Centre of Competence in Research, funded by the Swiss National Science Foundation (Grant number 205602). P.M. is grateful to N.A. Spaldin for useful discussions and acknowledges funding from the Swiss National Science Foundation (Grant number 209454). Calculations were performed on the \enquote{Euler} cluster of ETH Z\"urich. 
The Flatiron Institute is a division of the Simons Foundation.
\end{acknowledgments}

\appendix

\section{Atomic orbital-based projectors}
\label{app:NOA_and_LOAO}

QE offers the possibility to use both nonorthogonalized atomic orbitals (NAO) and LOAOs as Hubbard projectors (in addition to a few other options). NAO projector functions are simply atomic orbitals $\phi^I_{m \mathbf{R}}(\mathbf{r})$ that are provided with the pseudopotentials. They are orthonormal within each atom but not between different atoms. Here, $I$ is the atomic site index, $m$ is the magnetic quantum number associated with a selected orbital quantum number (representing the Hubbard manifold), and $\mathbf{R}$ is the Bravais lattice vector. However, in general it is more desirable to work with an orthonormal basis which, among other things, avoids counting Hubbard corrections twice in the interstitial regions between atoms. Therefore, LOAO are obtained by taking the atomic orbitals of each atom and orthonormalizing them to all orbitals of all other atoms in the system using the L\"owdin orthogonalization method~\cite{Lowdin:1950, Mayer:2002}. The Bloch sums of NAO are defined as:
\begin{equation}
|\phi^I_{\mathbf{k}m} \rangle = \sum_\mathbf{R} e^{i \mathbf{k \cdot R}} |\phi^I_{m \mathbf{R}} \rangle \,.
\label{eq:atomic}
\end{equation}
The Bloch sums of LOAO are in turn defined as:
\begin{equation}
    |\tilde{\phi}^I_{\mathbf{k}m} \rangle = \sum_{J m'} \left(O_\mathbf{k}^{-\frac{1}{2}}\right)^{JI}_{m' m} |\phi^J_{\mathbf{k}m'} \rangle \,,
    \label{eq:LOAO_def}
\end{equation}
where $O_\mathbf{k}$ is the orbital overlap matrix which is defined as:
\begin{equation}
    (O_\mathbf{k})^{IJ}_{m m'} = \braket{\phi^I_{\mathbf{k}m}}{\phi^J_{\mathbf{k}m'}} \,,
    \label{eq:LOAO_overlap_def}
\end{equation}

\section{Wannier functions from atomic projections}
\label{app:WFs}

As noted in \pref{subsec:wannier_projectors} and \pref{subsec:Comp_det}, in this work we refrain from minimizing the quadratic spread of our WFs and instead use the ``initial guesses'' that are obtained by Wannier90 based on projections of the Bloch functions on atomic orbitals and subsequent L\"owdin orthonormalization.
In this case, the Bloch sums of L\"owdin-orthonormalized WFs are defined as~\cite{Mostofi_et_al:2014}:
\begin{equation}
| \mathrm{w}^I_{m \mathbf{k}} \rangle = \sum_{Jm'} \left(O_\mathbf{k}^{-\frac{1}{2}}\right)^{JI}_{m'm} \, | \tilde{\mathrm{w}}^J_{m' \mathbf{k}} \rangle ,
\label{eq:WF_poor_man}
\end{equation}
where $O_\mathbf{k}$ is the overlap matrix defined as:
\begin{equation}
\left( O_\mathbf{k} \right)^{IJ}_{mm'} = \langle \tilde{\mathrm{w}}^I_{m \mathbf{k}} | \tilde{\mathrm{w}}^J_{m' \mathbf{k}} \rangle \,.
\label{eq:Overlap_matrix}
\end{equation}
Here, $|\tilde{\mathrm{w}}^I_{m \mathbf{k}} \rangle$ are the Bloch sums of non-orthogonalized WFs defined as:
\begin{equation}
|\tilde{\mathrm{w}}^I_{m \mathbf{k}} \rangle = \sum_{\nu} |\psi_{\mathbf{k} \nu} \rangle A^I_{\nu m \mathbf{k}} \,,
\label{eq:WF_non-ortho}
\end{equation}
where
\begin{equation}
A^I_{\nu m \mathbf{k}} = \langle \psi_{\mathbf{k} \nu} | \phi^I_{\mathbf{k}m} \rangle \,,
\label{eq:A_def}
\end{equation}
and $| \phi^I_{\mathbf{k}m} \rangle$ is defined in Eq.~\eqref{eq:atomic}. 

It is worth noting that equivalent WFs as in Eq.~\eqref{eq:WF_poor_man} have already been available in QE. They can be generated by the program \texttt{pmw.x}~\cite{Fabris:2005}, which is part of the official QE distribution.
The potential advantage of our new interface to the Wannier90 code is that it allows for more flexibility, since now any type of WF generated by Wannier90 can be used as Hubbard projector, including maximally localized ones~\cite{Marzari_et_al:2012}, symmetry-adapted ones~\cite{Sakuma:2013}, strictly atom-centered ones~\cite{Wang_et_al:2014}, or others.

\section{\label{app:Interface} Interface between Wannier90 and DFT+$U$ in Quantum ESPRESSO}

Using WFs from the Wannier90 code as Hubbard projectors in DFT+$U$ calculations with QE requires the following steps: 

$(i)$ An initial DFT+$U$ calculation (using either NAO or LOAO projectors) with the \texttt{pw.x} executable \IT{of QE}. Here, even if one wants use a plain DFT calculation as starting point, it is necessary to enable the DFT+$U$ machinery by specifying a negligibly small $U$ value, so that the required dimensions of the Hubbard projectors are written to the extensible markup language (XML) file.

$(ii)$ Construct WFs based on this initial calculation using the Wannier90 code~\cite{Pizzi2020} \response{and the \texttt{pw2wannier90.x} executable of QE}. Detailed instructions for this step are available in the Wannier90 documentation.
Here, it is crucial to set \texttt{write\_u\_matrices = .true.} in the Wannier90 input file, so that the gauge matrices $U_{\nu m\mathbf{k}}$ are written to the file \texttt{*\_u.mat} (and potentially \texttt{*\_u\_dis.mat}).

$(iii)$ Read the WFs from the Wannier90 output and use them to replace the Hubbard projectors from the initial DFT+$U$ calculation. This step uses our interface \response{in the \texttt{wannier2pw.x} executable of QE}, as described below. 

$(iv)$~Perform the DFT+$U$ production calculations using the constructed WFs as Hubbard projectors. In the input file for \texttt{pw.x}, the user must specify that the Hubbard projectors are ``\texttt{wf}'' in the HUBBARD card. 

We now describe the interface between Wannier90 and DFT+$U$ in QE, referenced in step $(iii)$ of the procedure above. This interface is implemented as a new, separate Fortran \response{\texttt{wannier2pw.x}} program of QE. It reads the WFs generated by Wannier90 (in form of the gauge matrices $U_{\nu m\mathbf{k}}$) and then replaces the NAO or LOAO Hubbard projectors with the selected WFs by writing them to the \texttt{.hub} files in the QE temporary folder.
To enable this, one must set the keyword \texttt{hubbard = .true.} in the input namelist of \response{\texttt{wannier2pw.x}}. 
In addition, the specific WFs that will be used as Hubbard projectors need to be specified using the \texttt{wan2hub(i)=.true.} keyword, where \texttt{i} is the index of the selected WF in the list generated by Wannier90 in step $(ii)$.
The selected WFs are read in the order they appear in the Wannier90 output and then successively replace the projectors from the initial calculation from step $(i)$. Their order therefore needs to be compatible with that of the original projectors in terms of site, orbital, and spin character.
If lower-energy states have been excluded from the Wannierization window, then also the keyword \texttt{exclude\_ks\_bands} needs to be specified, which sets the starting index for the summation over bands in \pref{eqn:Wannier_generic} to \texttt{exclude\_ks\_bands}+1.
Otherwise, the summation in \pref{eqn:Wannier_generic} starts from the lowest-energy KS state. 
The upper bound for the band summation is determined automatically from the data files (\texttt{*\_u.mat} or \texttt{*\_u\_dis.mat}) generated by Wannier90. 

The interface supports WFs obtained for both entangled and disentangled KS bands. Additionally, the interface supports norm-conserving and ultrasoft pseudopotentials~\cite{Vanderbilt:1990}, as well as the PAW method~\cite{Blochl:1994}. 
Since symmetry is currently not implemented in Wannier90, the full $\mathbf{k}$-point grid in the first Brillouin zone must be used. Therefore, the user must set \texttt{nosym=.true.} and \texttt{noinv=.true.} in the \texttt{pw.x} input file.

Finally, we point out that in general the values for $U$ (and $J$) must match the definition of the Hubbard projectors, and values obtained for a specific choice of projectors might not be suitable if used with another choice of projectors.

\bibliography{main}
\end{document}


\title{Supplemental materials to the manuscript ``Explicit demonstration of the equivalence between DFT+$U$ and the Hartree-Fock limit of DFT+DMFT''}

\author{Alberto Carta} \email{alberto.carta@mat.ethz.ch}
\affiliation{Materials Theory, ETH Z\"urich, Wolfgang-Pauli-Strasse 27, 8093 Z\"urich, Switzerland}
\author{Iurii Timrov} \email{ iurii.timrov@psi.ch}
\affiliation{PSI Center for Scientific Computing,
Theory, and Data, 5232 Villigen PSI, Switzerland}
\author{Peter Mlkvik}
\affiliation{Materials Theory, ETH Z\"urich, Wolfgang-Pauli-Strasse 27, 8093 Z\"urich, Switzerland}
\author{Alexander Hampel}
\affiliation{Center for Computational Quantum Physics, Flatiron Institute, 162 5th Avenue, New York, NY 10010, USA}
\author{Claude Ederer} \email{edererc@ethz.ch}
\affiliation{Materials Theory, ETH Z\"urich, Wolfgang-Pauli-Strasse 27, 8093 Z\"urich, Switzerland}

\date{\today}

\maketitle

\section{Additional computational details}


\subsection{NiO and MnO}

For NiO and MnO we perform calculations using a unit cell containing 2 formula units which allows us to describe the experimentally observed antiferromagnetic order with wave vector along the [111] crystallographic direction. The rhombohedral distortions of the lattice are neglected.
We fully relax the crystal structure for the non-magnetic, i.e., non-spin-polarized case, using the local-density approximation (LDA)~\cite{Perdew1981} with ultrasoft pseudopotentials from the GBRV library~\cite{Garrity2014_GBRV_pseudos}. As lattice constants of the conventional cell, we obtain 4.18 \AA \, for NiO and 4.31 \AA \, for MnO.
For both materials we employ a $6 \times 6 \times 6 $ $\mathbf{k}$-point grid with the kinetic-energy plane-wave cutoff of 50 and 500~Ry for the Kohn-Sham wavefunctions and density, respectively.

\begin{figure}
   \centering   \includegraphics[width=0.8\textwidth]{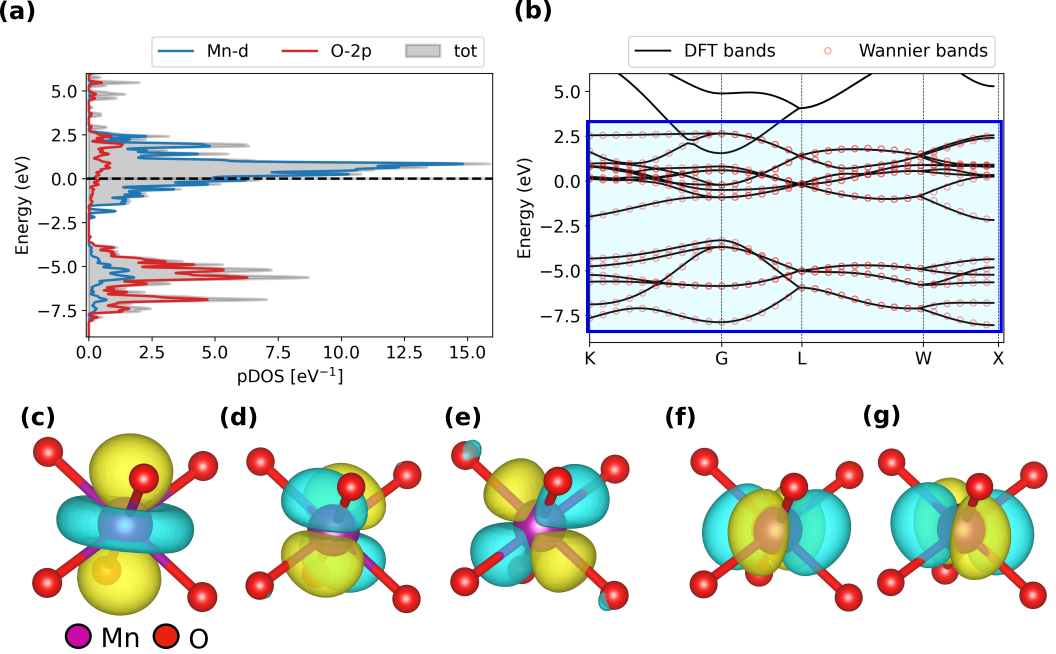}
   \caption{(a) MnO density of states together with the projections  on the Mn $3d$ (blue line) and O $2p$ (red line) states. The zero of energy corresponds to the Fermi level. (b) MnO band structure obtained from spin-unpolarized DFT calculations (black lines) together with bands reconstructed from the tight-binding Hamiltonian expressed in the WFs basis (red circles). The blue rectangle reflects the Wannier energy window which includes all states with majority O $2p$ and Mn $3d$ character. (c)-(g) Plots of the corresponding $d$-like WFs centered on the Mn atom.}
   \label{fig:mno_big}
\end{figure}

Figure~\ref{fig:mno_big} shows the calculated projected and total densities of states as well as the band structure of MnO. Analogous to Fig.~1 in the main text for NiO, the energy window used for the construction of the Wannier functions and plots of the resulting Mn-centered $d$-like Wannier functions are also shown here for MnO.


\subsection{\lamno}

For calculations on \lamno, we take the experimentally determined structure with space group $Pbnm$ as determined by neutron diffraction at room temperature~\cite{RodrguezCarvajal1998}.
We consider a $\sqrt{2} \times \sqrt{2} \times 2$ unit cell containing 4 formula units, for which we employ a $7 \times 7 \times 5$ $\mathbf{k}$-point grid.
We use the PBE exchange-correlation functional~\cite{Perdew1996_PBE} together with ultrasoft pseudopotentials from the \QE pseudopotential library: \url{https://pseudopotentials.quantum-espresso.org/legacy_tables}. We use the kinetic-energy plane-wave cutoff of 60 and 720~Ry for the Kohn-Sham wavefunctions and density, respectively.
Experimentally, the material exhibits an A-type antiferromagnetic order (wave vector along [001])~\cite{RodrguezCarvajal1998,Wollan1955} which is what we impose in our calculations.

\begin{figure}
   \centering   \includegraphics[width=0.9\textwidth]{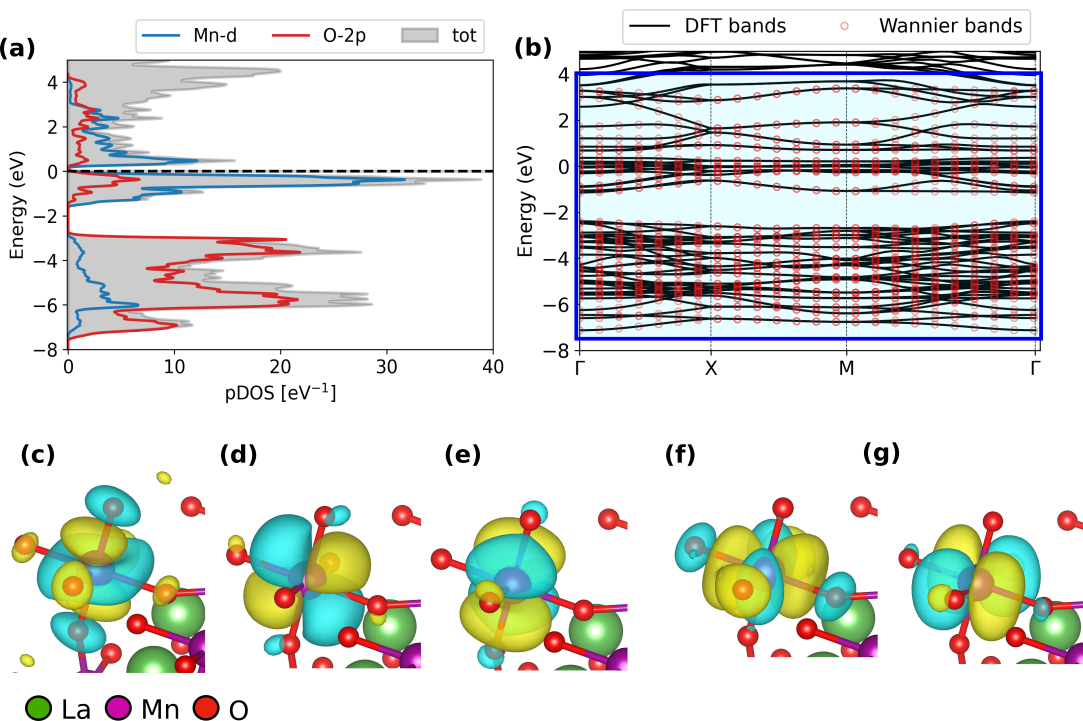}
   \caption{(a) \lamno density of states together with the projection on the Mn $3d$ (blue line) and O $2p$ (red line) states. The zero of energy corresponds to the Fermi level. (b)~\lamno band structure obtained from spin-unpolarized DFT (black lines) together with bands reconstructed from the Wannierization (red circles). The blue rectangle represents the Wannier energy window which includes all states with majority O $2p$ and Mn $3d$ character. (c)-(g) Plots of the $d$-like Wannier functions centered on the Mn atom.}
   \label{fig:lamno_big}
\end{figure}

Figure~\ref{fig:lamno_big} depicts the projected and total density of states, the DFT and Wannierized band structures, the energy window used to construct the Wannier functions, and plots of the resulting $d$-like Wannier functions centered on the Mn atoms. 


\subsection{\lunio}

We consider a $\sqrt{2} \times \sqrt{2} \times 2$ unit cell containing 4 formula units of \lunio, which we relax using spin-unpolarized PBE as the exchange-correlation functional \cite{Perdew1996_PBE} together with ultrasoft pseudopotentials from the \QE pseudopotential library: \url{https://pseudopotentials.quantum-espresso.org/legacy_tables}.
We use a $\mathbf{k}$-point grid of size $7 \times 7 \times 5$.
We use the kinetic-energy plane-wave cutoff of 60 and 720~Ry for the Kohn-Sham wavefunctions and density, respectively.
We begin by relaxing the structure for a spin-unpolarized calculation, which resulted in crystal structure with space group $Pbnm$. 
Following previous studies \cite{Hampel2017, Hampel2020}, we quantify the degree of Ni-O bond disproportionation using the $R_1^+$ mode obtained through symmetry mode decomposition using ISODISTORT \cite{Campbell2006}.
To impose a local spin-splitting on the Ni atoms, we consider an A-type magnetic order~\cite{Hampel2017}, despite it not being the magnetic ground state of the system, which is instead a more complex arrangement of the Ni moments~\cite{SerranoSanchez2022}.

\begin{figure}
   \centering   \includegraphics[width=0.9\textwidth]{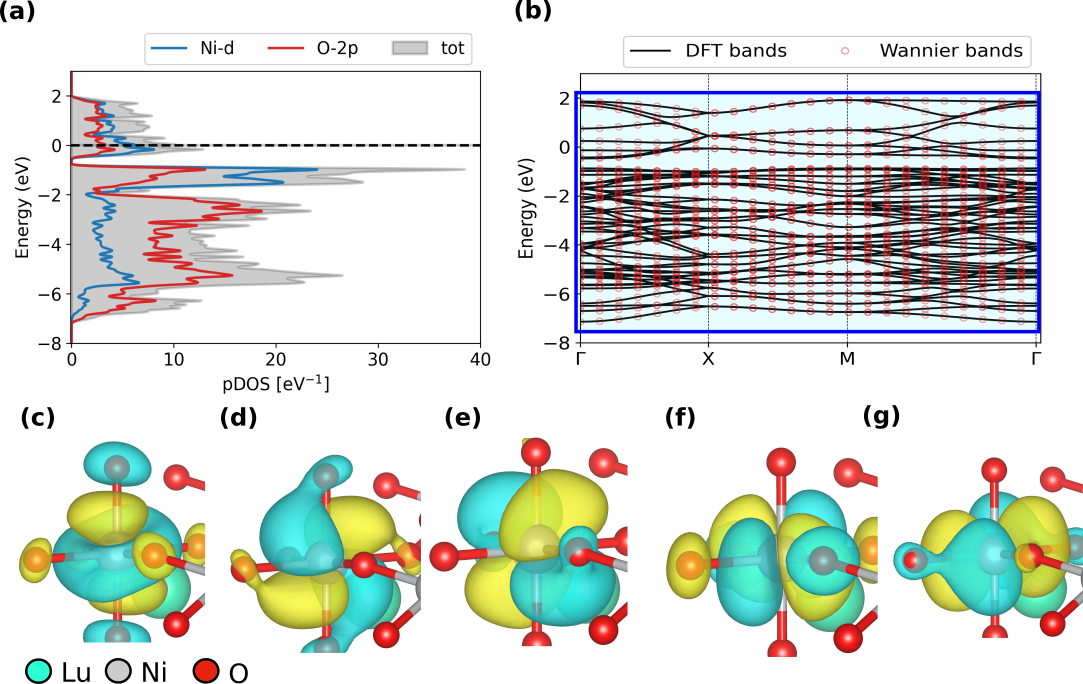}
   \caption{(a) Density of states of the spin-unpolarized $Pbnm$ relaxed \lunio structure. The projection on the Ni $3d$ states is indicated in blue and on the O $2p$ states in red. The zero of energy corresponds to the Fermi level. (b) \lunio band structure obtained from spin-unpolarized DFT (black lines) together with bands reconstructed from the Wannierization (red circles). The blue rectangle represents the Wannier energy window which includes all states with majority O $2p$ and Ni $3d$ character. (c)-(g) Plots of the $d$-like Wannier functions centered on the Ni atom.}
   \label{fig:lunio_big}
\end{figure}

Figure~\ref{fig:lunio_big} depicts the projected and total density of states, the DFT and Wannierized band structures, the energy window used to construct the Wannier functions, and plots of the resulting $d$-like Wannier functions centered on the Ni atoms. 


\subsection{VO$_2$}

We consider a monoclinic cell encompassing 4 formula units of VO$_2$, constructed from the lattice vectors of the experimental rutile (R) structure~\cite{McWhan_et_al:1974}. In our treatment of the monoclinic (M1) phase, for simplicity, we disregard the experimentally observed unit cell expansion along $c$ and also all other strain components relative to the R phase. To scan over the distortion between R and M1 phases and beyond, we linearly interpolate and extrapolate the internal atomic positions while keeping the lattice vectors fixed.

We employ the spin-unpolarized PBE exchange-correlation functional and the ultrasoft pseudopotentials from the GBRV library~\cite{Garrity2014_GBRV_pseudos} with semicore 3$s$ and 3$p$ states included as valence for the V~atoms. We use a $\mathbf{k}$-point grid of size $6 \times 6 \times 8$.
We use the kinetic-energy plane-wave cutoff of 70 and 840~Ry for the Kohn-Sham wavefunctions and density, respectively.

\begin{figure}
   \centering  \includegraphics[width=0.9\textwidth]{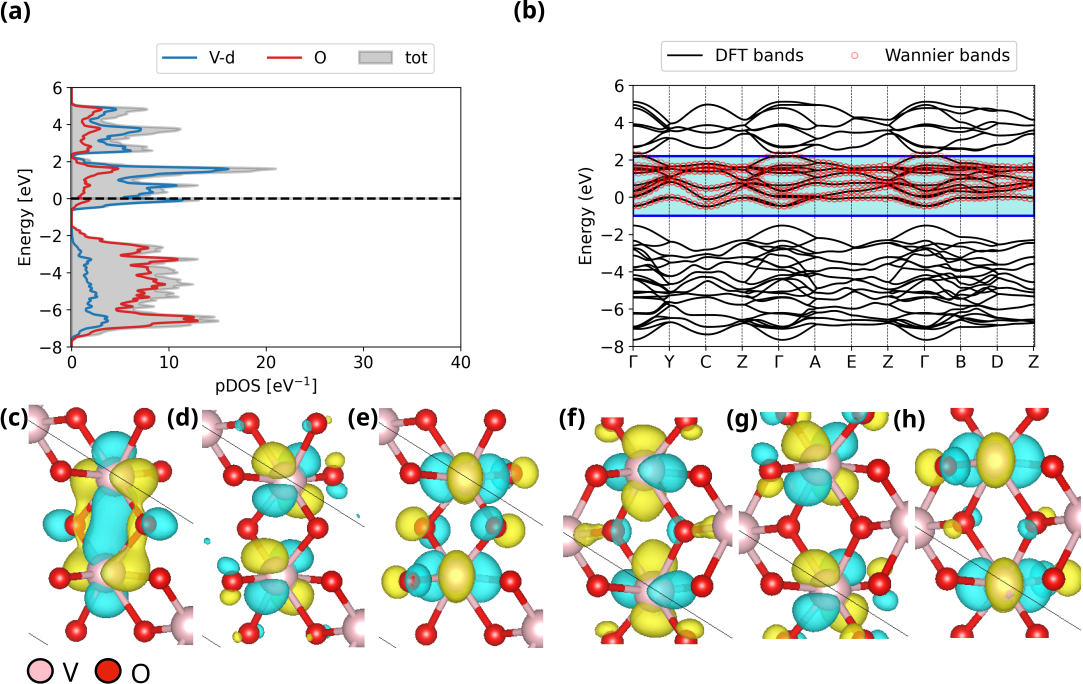}
   \caption{(a) Density of state of the M1 VO$_2$ structure. The projection on the V $3d$ states is indicated in blue and on the O $2p$ states in red. The zero of energy corresponds to the Fermi level. (b) VO$_2$ band structure obtained from spin-unpolarized DFT (black lines) together with bands reconstructed from the Wannierization (red circles). The blue rectangle represents the Wannier energy window which includes all the V $d$ majority $t_{2g}$ states. (c)-(h) Plots of the Wannier functions centered on the two inequivalent V-V bonds.}
   \label{fig:vo2_big}
\end{figure}

To construct the bond-centered Wannier functions, we follow the procedure according to Ref.~\cite{Mlkvik_et_al:2024}. We use a frozen energy window which encompasses most of the $t_{2g}$-dominated band manifold to first construct conventional atom-centered Wannier functions and then define a bond-centered set by a pair-wise unitary transformation, resulting in 3 Wannier functions per bond [see Fig.~\ref{fig:vo2_big}~(c)-(h) for the two inequivalent sets of short-bond and long-bond Wannier functions].

We describe Coulomb interaction on each bond-centered Wannier functions by a Hubbard-Kanamori Hamiltonian~\cite{Vaugier/Jiang/Biermann:2012} including the spin-flip and pair-hopping terms, with $\tilde{U}$ corresponding to the intra-orbital Coulomb interaction and $\tilde{J}$ the Hund's coupling. For more details see Ref.~\cite{Mlkvik_et_al:2024}.

Within our DFT+DMFT(HF) calculation, we obtain the local occupations on a given site (bond center) as $n_{mm^\prime} = G_{mm^\prime}(\tau = 0^{-})$ where $G_{mm^\prime}(\tau)$  is the local Green's function for the different orbitals $m$ and $m^\prime$ at the imaginary time $\tau$. We also report the averaged spectral weight around the Fermi level as $\bar{A}(\omega=0)=-(\frac{\beta}{\pi})$Tr$G(\tau=\frac{\beta}{2})$, where $\beta = (k_BT)^{-1}$ is the inverse electronic temperature.

Figure~\ref{fig:vo2_big} depicts the projected and total density of states, the DFT and Wannierized band structures, the energy window used to construct the Wannier functions, and plots of the resulting $d$-like Wannier functions centered on the bonds between the V atoms.

\bibliography{main}